\documentclass[reprint,amsmath,amssymb,aps,floatfix,prd,altaffilletter,nofootinbib]{revtex4-2}

\usepackage{graphicx}
\usepackage{dcolumn}
\usepackage{bm}
\usepackage[T1]{fontenc}
\usepackage{subfigure}
\usepackage{indentfirst}
\usepackage{amsmath}
\usepackage{soul}
\usepackage[normalem]{ulem}
\usepackage{xcolor}
\usepackage{multirow}

\newcommand{\dd}{\mathrm{d}}
\newcommand{\Mpl}{M_\mathrm{Pl}}
\newcommand{\Mpc}{\mbox{Mpc}}
\newcommand{\Hz}{\mbox{Hz}}

\usepackage{hyperref}

\begin{document}

\preprint{APS/123-QED}

\title{Scalar-Induced Gravitational Waves from self-resonant preheating in \texorpdfstring{$\bm\alpha$}{alpha}-attractor models}

\author{Daniel del-Corral$^{1,2}$}
\email{corral.martinez@ubi.pt}

\author{Paolo Gondolo$^{3,4}$}
\email{paolo.gondolo@utah.edu}

\author{K. Sravan Kumar$^{5}$}
\email{sravan.kumar@port.ac.uk}

\author{João Marto$^{1,2}$}
\email{jmarto@ubi.pt}

\affiliation{$^{1}$ Departamento de Física, Universidade da Beira Interior, Rua Marquês D'Ávila e Bolama 6200-001 Covilhã, Portugal}
\affiliation{$^{2}$Centro de Matemática e Aplicações da Universidade da Beira Interior, Rua Marquês D'Ávila e Bolama 6200-001 Covilhã, Portugal}
\affiliation{$^{3}$Department of Physics and Astronomy, University of Utah, Salt Lake City, UT 84112, USA}
\affiliation{$^{4}$Department of Physics, Institute of Science Tokyo, 2-12-1 Ookayama, Meguro-ku, Tokyo 152-8551, Japan}
\affiliation{$^{5}$Institute of Cosmology and Gravitation, University of Portsmouth, Dennis Sciama Building, Burnaby Road, Portsmouth, PO1 3FX, United Kingdom}


\begin{abstract}
After the inflationary phase, the universe enters the preheating phase, during which the inflaton field rolls down its potential and oscillates. When the potential significantly deviates from a parabolic shape at its minimum, these oscillations trigger an instability in the scalar perturbations, leading to their amplification. This phenomenon, known as self-resonance, has important implications in cosmology. Notably, since scalar perturbations couple to tensor perturbations at second order in the equations of motion, this amplification results in the production of Gravitational Waves (GWs), referred to as Scalar-Induced Gravitational Waves (SIGWs). In this study, we investigate the production of SIGWs during the preheating phase for a class of inflationary models known as $\alpha$-attractors, characterized by a single parameter $\alpha$. We focus on small values of this parameter, specifically $\alpha \sim \mathcal{O}(10^{-1} - 10^{-4})$, where the self-resonance effect is particularly pronounced. We obtain lower bounds on this parameter, {$\log_{10}(\alpha)>-3.71 (-3.51)$} for the T-model and {$\log_{10}(\alpha)>-3.48 (-3.19)$} for the E-model, and assuming a duration of preheating of 1 e-folds (5 e-folds). These bounds are obtained based on the energy density of SIGWs constrained by Big Bang nucleosynthesis, which ultimately translates into lower bounds on the tensor-to-scalar ratio, {$r>6.54\times10^{-7} (1.03\times10^{-6})$ for the T-model and $r>1.11\times10^{-6} (2.14\times10^{-6})$} for the E-model. {Note that these bounds on $\alpha$ and $r$ are derived within the linear framework of tensor fluctuations at the level of equations of motion, which nevertheless include scalar-scalar-tensor interactions with metric and matter fields. However, fully non-linear approaches, with all higher-order metric fluctuations, would be needed in the future to further validate these conclusions.}
\end{abstract}

\maketitle



\section{Introduction}

After the Planck 2013-2018 data, there has been a surge in the building of single-field Starobinsky-like inflationary scenarios \cite{Linde:2014nna} in various frameworks of quantum gravity due to stringent constraints on multifield inflation. The vanilla models of single-field inflation predict the scalar spectral index ($n_s$) and tensor-to-scalar ratio ($r$) as \cite{Kehagias:2013mya,Ellis:2013nxa}
\begin{equation}\label{vpred}
    n_s\approx 1-\frac{2}{N},\quad r=\frac{12\alpha}{N^2}\,,
\end{equation}
where $\alpha$ is a parameter that depends on the framework for the ultra-violet (UV) completion of gravity. In the context of $\alpha$-attractor inflation, the $\alpha$ parameter indicates the curvature of the K\"{a}hler geometry in supergravity \cite{Kallosh:2013yoa,Kallosh:2013daa,Kallosh:2013hoa,Roest:2013fha,Kaiser:2013sna,Kallosh:2015lwa,Iacconi:2023mnw,Carrasco:2015uma,Kallosh:2015zsa}. It is important to note that models of $\alpha$-attractors could predict, in general, any smaller value for the tensor-to-scalar ratio compatible with the latest bound $r<0.032$ \cite{BICEP:2021xfz,Tristram:2021tvh}. Although future observations aim to detect primordial gravitational waves, there is a lack of a theoretical lower bound on the value of $\alpha$ except through considerations of reheating \cite{Iacconi:2023mnw}. However, right after inflation ends, the universe enters a phase known as preheating \cite{Dolgov:1989us,Traschen:1990sw,Shtanov:1994ce,Kofman:1994rk,Kofman:1997yn} during which the universe behaves effectively close to  matter-dominated phase with instabilities \cite{Martin:2019nuw,Martin:2020fgl,del-Corral:2023apl,del-Corral:2024vcm}. Furthermore, when the inflationary potential deviates significantly from quadratic behavior (as is the case for small $\alpha$), the oscillations trigger the phenomenon of self-resonance {(either by including or excluding metric fluctuations)} \cite{del-Corral:2023apl,del-Corral:2024vcm,Hertzberg:2014iza,Hertzberg:2014jza,Jedamzik:2010dq,Martin:2019nuw,Sfakianakis:2018lzf,Ballesteros:2024hhq,Martin:2020fgl,Shafi:2024jig,Mahbub:2023faw,Sang:2020kpd,Zhang:2023hjk,Child:2013ria,Sang:2019ndv,Jia:2024fmo}, which translates into a fast and substantial amplification of scalar perturbations that act as a source of Gravitational Waves (GWs), since these fluctuations couple to tensor ones at second order in the equations of motion or 3rd order in the perturbation of the action. The GWs produced in this fashion are called Scalar-Induced Gravitational Waves (SIGWs) \cite{Ananda:2006,Assadullahi:2009,Baumann:2007,Guzzetti:2016,Maggiore:2007,Domenech:2021,Kohri:2018awv,Franciolini:2021}, and can be so efficiently produced to reach the Big Bang Nucleosynthesis (BBN) bound. This precise fact is what our investigation utilizes to derive a new lower bound on $\alpha$, based solely on the considerations of preheating instabilities and the consequent generation of SIGWs. Our result is robust even for a small duration of preheating of 1-2 e-folds, as one can realistically expect after the end of inflation. 

Recently, we have shown that these instabilities during preheating could lead to the generation of primordial black holes (PBH) \cite{del-Corral:2023apl,draft-alpha-PBH} in Starobinsky as well as in $\alpha$-attractor inflationary models. However, the study in this paper focuses on SIGWs produced irrespective of whether or not PBH are generated during preheating. It is also important to note that SIGWs are a secondary effect. At the linearized level, both scalar and tensor perturbations decouple. Therefore, we compute the linearized effect, which we call background GWs (BGWs), and the non-linear (secondary) effect, \textit{i.e.}, the SIGWs. 
\begin{figure}
    \centering
    \includegraphics[trim = 20mm 110mm 5mm 0mm,width=\linewidth]{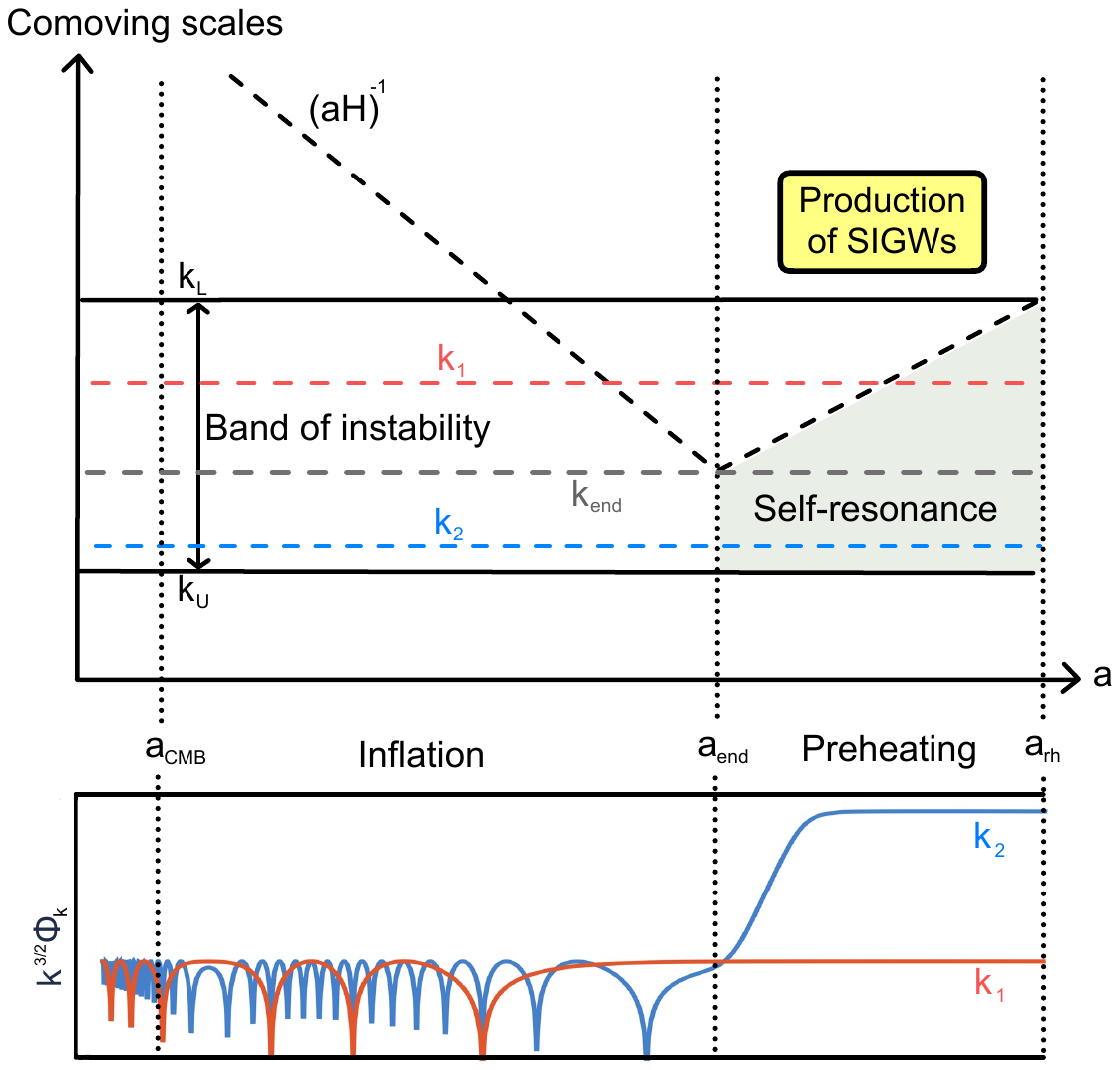}
    \caption{Schematic representation of the production of SIGWs during self-resonant preheating. The upper panel shows the evolution of selected comoving scales as a function of the scale factor of the universe, as well as the comoving Hubble radius $(aH)^{-1}$. The suffixes \textit{CMB}, \textit{end}, and \textit{rh} correspond to the CMB pivot scale, the end of inflation, and the beginning of reheating, respectively. The evolution of {the Fourier mode of the metric fluctuation} $\Phi_k$ in the bottom panel is shown schematically to illustrate the approximate behavior of the perturbations. See the text for details.}
    \label{fig:scheme}
\end{figure}
A schematic representation of the scenario described above is shown in Fig.~\ref{fig:scheme}. The key aspect is the definition of the relevant scales involved in the instability, labeled as \textit{band of instability}. This band is bounded by two characteristic scales: $k_{L}$ (lower) and $k_{U}$ (upper). The lower bound, $k_{L}$, corresponds to the last mode entering the horizon during preheating and is defined as $k_{L} = a(t_{\text{rh}})H(t_{\text{rh}})$, where $a(t)$ is the scale factor of the universe, $H(t)$ the Hubble rate, and $t_{\text{rh}}$ marks the end of the preheating phase. The upper bound, $k_{U}$, denotes the largest mode affected by self-resonance and is determined using Floquet theory~\cite{del-Corral:2024vcm,Martin:2020fgl}. When the modes enter the self-resonance instability (green shaded area), the scalar perturbations, such as $\Phi_{\bm k}$, amplify. Their behavior is illustrated in the bottom panel of Fig.~\ref{fig:scheme} for two distinct scales, $k_1$ and $k_2$, separated by the scale $k_\text{end}=a_\text{end}H_\text{end}$, defined as the last scale to exit the horizon during inflation {(here and in what follows, the suffix "\textit{end}" refers to evaluation at the end of inflation)}. The first {scale}, shown in red, remains approximately constant after re-entering the Hubble horizon $(aH)^{-1}$ during preheating. Although this may appear as non-amplification, it actually reflects the counteraction of the usual decay of the gravitational potential at sub-horizon scales in an expanding universe, ensuring $\Phi_{\bm{k}}$ remains stable during this phase. The mode $k_2$, shown in blue, experiences stronger amplification as it spends more time within the region of instability of self-resonance. It should be noted that the Floquet exponent, responsible for amplification, is not the same for all modes within the band of instability \cite{del-Corral:2024vcm}. In fact, the amplification is particularly stronger for modes $k>k_\text{end}$. During this amplification process, SIGWs are generated, and after that, the SIGWs freely propagate throughout the universe with minimal interaction with matter.

Depending on the specific range of scales affected by the self-resonance, the SIGWs will have an imprint in a particular span of frequencies. In general, these scales are small and around the last scale to exit the horizon during inflation, which implies that we lie on the Very-High-Frequency (VHF) region of the GW spectrum (100 kHz -- 1 THz) \cite{Kuroda:2015owv,Aggarwal:2020olq,Page:2020zbr,Dandoy:2024oqg}. This VHF region has been explored in the literature in the contexts of GUT scale PBH and quantum gravity \cite{Anantua:2008am,Zagorac:2019ekv,Addazi:2024qcv}.
Although some detectors (in operation and/or planned) work on the VHF band of the GW spectrum, their sensitivities are too low to detect a stochastic background of SIGWs. These sensitivities are shown in Fig.~\ref{fig:sensitivity-curves} in terms of the fractional energy density of gravitational waves, $\Omega_{\text{GW}}$. One can observe that all of them lie above the cosmological bound coming from Big Bang nucleosynthesis (BBN) \cite{Smith:2006nka,Maggiore:1999vm}, making it impossible to detect a stochastic background of GWs formed during preheating (otherwise, it would conflict with BBN predictions). In the VHF band, one of interest in this work, the most promising probes to detect GWs are: 
\begin{figure*}
    \centering
    \includegraphics[width=0.75\linewidth]{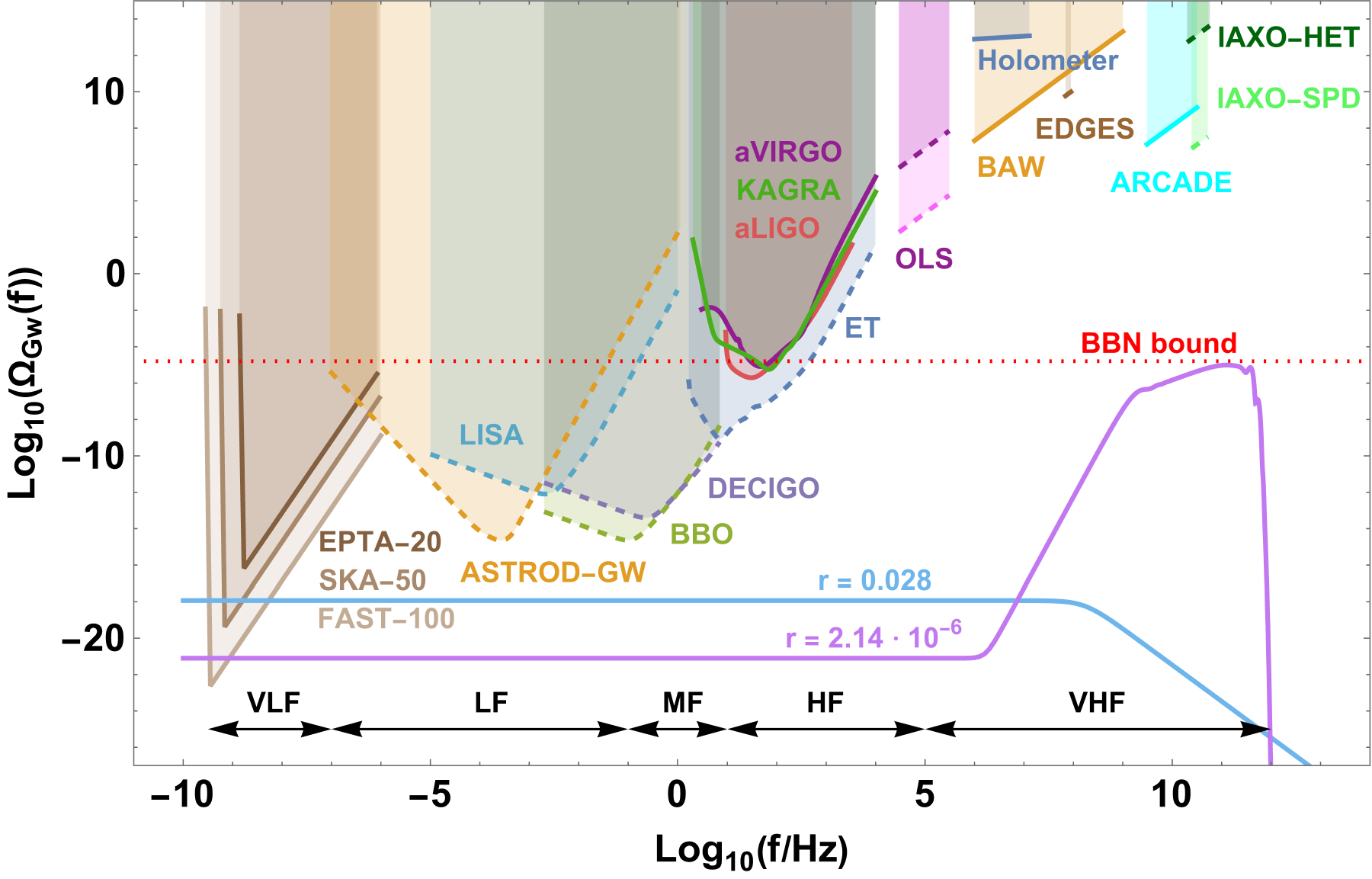}
    \caption{Sensitivity curves of some GW detectors (in terms of $\Omega_{GW}$) as a function of the frequency in Hertz. Also shown is the BBN bound, taken from \cite{Smith:2006nka,Maggiore:1999vm}. Solid (dashed) lines represent operative (planned or theorized) GW detectors. The figure is produced from the analysis of \cite{Kuroda:2015owv,Aggarwal:2020olq} and references therein. The bottom lines with arrows mark the limits of the standard frequency bands of GWs, as defined in \cite{Kuroda:2015owv}. The blue and magenta curves labeled as $r=0.028$ and $r=2.14\times10^{-6}$ represent two spectra of GWs computed from the upper and lower bounds on the tensor-to-scalar ratio; see the text for details. These are computed using the method outlined in this work.}
    \label{fig:sensitivity-curves}
\end{figure*}
\begin{itemize}
    \item \textit{Optically levitated sensors (OLS)}: These are based on the use of optically trapped and cooled dielectric microspheres or microdisks that experience a force when a GW passes through \cite{Aggarwal:2020umq}. 
    
    \item \textit{Bulk acoustic wave (BAW) devices}: The vibrations of a resonant mass detector due to the passage of a GW could be read through {specific devices  \cite{Goryachev:2014yra,Goryachev:2021zzn}.} 
    
    \item \textit{Interferometers up to 100 MHz}:
    {Laser interferometers measure differential arm-length changes induced by GWs.}
    The quantum Cramér-Rao bound implies that, for a given laser power, higher bandwidth is needed to increase the sensitivity \cite{Mizuno:1995iqa}. This implies that a broadband interferometer with the Virgo, LIGO, or KAGRA-level sensitivity is not viable in the MHz region. However, there are several other interferometers in this bandwidth, such as the Holometer experiment at Fermilab \cite{holometer}. 
    
    \item \textit{Gertsenshtein effect}: This consists of the conversion of photons into GWs (and vice versa) in the presence of a magnetic field. Although these kinds of detectors are not yet built, Ref.~\cite {Aggarwal:2020olq} points out the possibility of using existing experiments, such as IAXO{-SPD and IAXO-HET \cite{Ringwald:2020ist}}. Another possibility is to rely on observations of large-scale cosmic regions with radio telescopes, such as EDGES or ARCADE \cite{Domcke:2020yzq,Lella:2024dus}.
\end{itemize}

Apart from the VHF band, we also show for comparison in Fig.~\ref{fig:sensitivity-curves} the sensitivity curves of GW detectors in lower frequency bands: the Very-Low-Frequency (VLF), the Low-Frequency (LF), the Middle-Frequency (MF), and the High-Frequency (HF). See \cite{Kuroda:2015owv,Aggarwal:2020umq} and references therein for details about the detectors and the acronyms. The two curves labeled as $r=0.028$ and $r=2.25\times10^{-6}$ in Fig.~\ref{fig:sensitivity-curves} correspond to the energy density of GWs (background $+$ induced) from an E-model with $\alpha=8.4$ and $\alpha=10^{-3.17}$, respectively, where $r$ is the tensor to scalar ratio \eqref{vpred}. The upper bound on $r$ corresponds to the one obtained using 10 datasets from the BICEP/Keck Array 2015
and 2018, Planck releases 3 and 4, and LIGO-Virgo-KAGRA Collaboration \cite{Galloni:2022mok}. The lower bound is obtained in this work using SIGWs and the BBN bound. See Sec.~\ref{sec:numerical-strategy} for details.

\begin{figure}[t]
    \centering
    \subfigure[T-model]{%
    \includegraphics[width=0.75\linewidth]{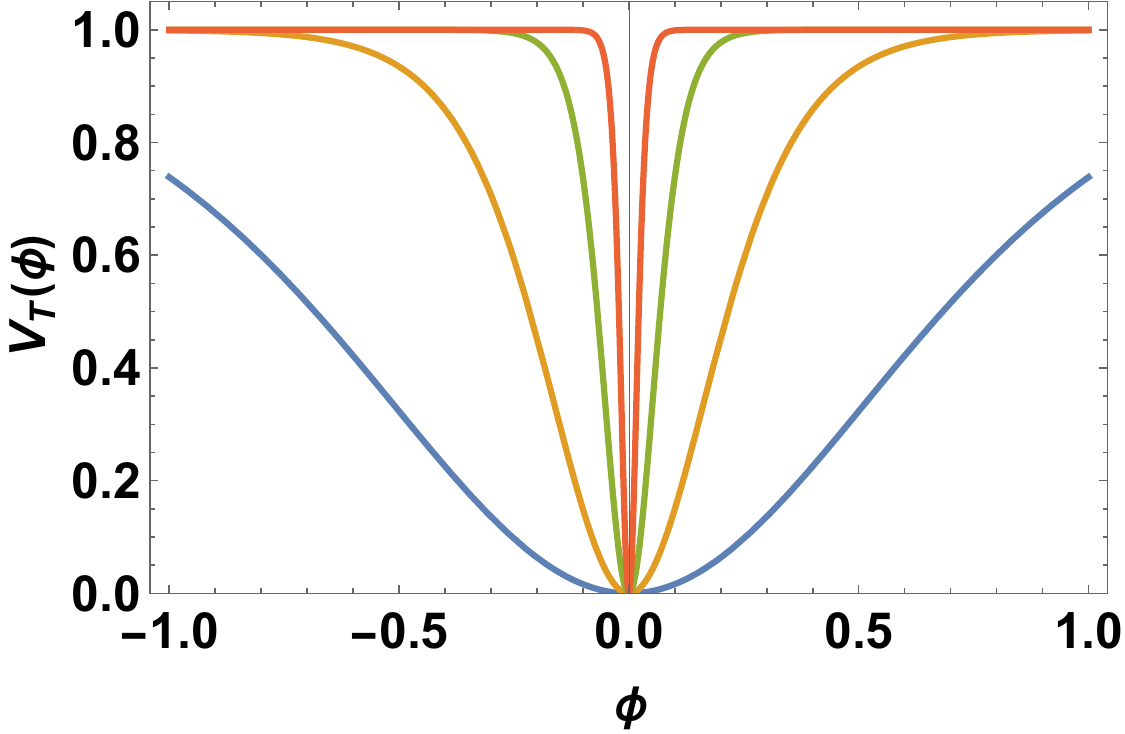}
        \label{fig:potentialsT}
    }
    \hfill
    \subfigure[E-model]{%
        \includegraphics[width=0.75\linewidth]{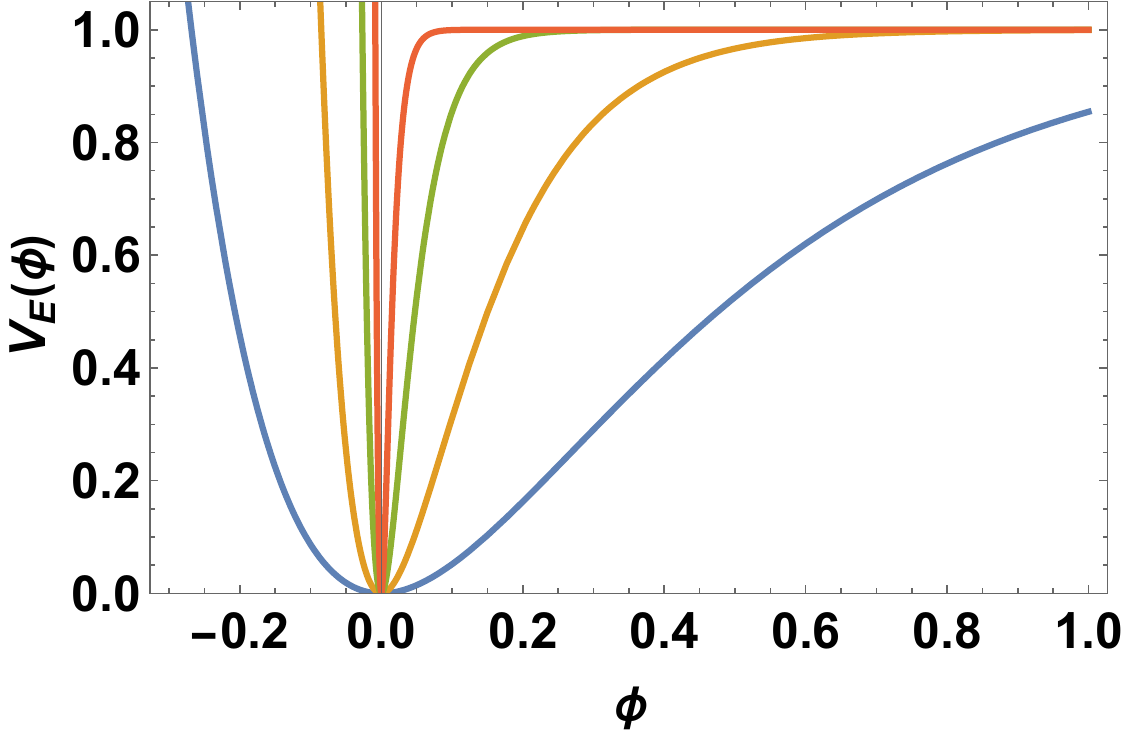}
        \label{fig:potentialsE}
    }
    \caption{Normalized $\alpha$-attractor potentials for several values of $\alpha$: $10^{-1}$ (blue), $10^{-2}$ (orange), $10^{-3}$ (green), and $10^{-4}$ (red).}
    \label{fig:potentials}
\end{figure}

The inflationary models we study in this work are {one of} the well-known {universality-class of inflationary potentials \cite{Roest:2013fha}, such as the }$\alpha$-attractor models \cite{Kallosh:2013yoa,Kallosh:2013daa,Kallosh:2013hoa,Kaiser:2013sna,Kallosh:2015lwa,Iacconi:2023mnw,Carrasco:2015uma,Kallosh:2015zsa}, which exhibit excellent agreement with Planck data \cite{Planck:2018jri} and have universal predictions of observables. For this reason, they have gained a lot of attention. There are two main subclasses of $\alpha$-attractor models, called T-models and E-models, named by the shape of the inflaton potentials (see Fig.~\ref{fig:potentials}), which are given by
\begin{equation}\label{eq:T-model}
\begin{split}
    V_T(\phi)&=3\alpha M^2\Mpl^2\tanh^2\left[\frac{1}{\sqrt{6\alpha}}\phi\right],\\ V_E(\phi)&=\frac{3\alpha M^2\Mpl^2}4\left(1-e^{-\sqrt{\frac{2}{3\alpha}}\phi}\right)^2,
\end{split}
\end{equation}
where $\Mpl$ is the Planck mass and {the scalar field mass} $M= 1.3\times 10^{-5} \Mpl$ is fixed by the amplitude of primordial power spectra. Beyond their inflationary observables \eqref{vpred}, we show that these models are also particularly interesting from the perspective of SIGWs. As the value of $\alpha$ decreases, the minimum of the potential (at $\phi=0$) starts to deviate from the quadratic (parabola) behavior, triggering the self-resonance phenomenon and amplifying the scalar perturbations. {Here we restrict our study to \eqref{eq:T-model} potentials. 
Fig.~\ref{fig:potentials} compares both potentials {in Eqn.~\eqref{eq:T-model}} for several values of $\alpha$, revealing an evident asymmetry in the E-model case. This, as shown in \cite{del-Corral:2024vcm}, makes the self-resonance effect stronger in this class of potentials, ultimately imposing a tighter constraint on $\alpha$ for the E-models and, consequently, on the tensor-to-scalar ratio. These differences will be further analyzed in Sec.~\ref{sec:tensor-perturbations}, where their impact on tensor perturbations is clarified.}

{It would be non-trivial to extend our results to higher order $T, E-$ shaped potentials of universality class \cite{Roest:2013fha}. For example, in the case of a potential with a quartic minimum (that comes from potential class $V_T \sim \tanh^4\left( \frac{\phi}{\sqrt{6\alpha}}\right)$ \cite{Roest:2013fha}), the perturbations do amplify, as shown in \cite{Kofman:1994rk,Kaiser:1997hg,Bassett:2005xm,Jedamzik:2010dq}. However, as a consequence of this, the resonance band is likely to get narrowed compared to the potentials in \eqref{eq:T-model}. Although we do not show the results for this class of potentials, we expect this to affect the production of SIGWs in two ways. First, the universe in this case behaves effectively as radiation-dominated right after inflation, affecting the production of SIGWs, and second, this production, if any, would correspond to a narrower range of frequencies in comparison, but could also result in a stronger spectrum of GWs depending on the details of inflaton dynamics prior to its decay. The quantification of this effect would require further investigation, which goes beyond the scope of the paper. }

This work is organized as follows: In Sec.~\ref{sec:scalar-perturbations}, we analyze the evolution of scalar metric perturbations in a perturbed flat Friedmann-Lema\^itre-Robertson-Walker (FLRW) metric. In Sec.~\ref{sec:tensor-perturbations}, we examine the generation of SIGWs from these scalar perturbations and numerically compute the energy density of GWs for different values of $\alpha$. Finally, conclusions are drawn in Sec .~\ref {sec:conclusions}, and Appendixes~\ref{sec:EOM-SIGWs} and \ref{sec:PS-SIGWs} show some complementary computations, such as the derivation of the equation of motion for the SIGWs or its power spectrum.
Throughout the paper, we follow the metric signature $(-+++)$ and $\hbar=c=1$ with reduced Planck mass $M_{\rm Pl}^2 =\frac{1}{8\pi G}$. Throughout the paper, overdots denote differentiation with respect to cosmic time ($t$) and overprimes denote differentiation with respect to conformal time ($\eta$). 


\section{Scalar perturbations}\label{sec:scalar-perturbations}

\begin{figure}[t]
    \centering
    \subfigure[]{\includegraphics[width=0.95\linewidth]{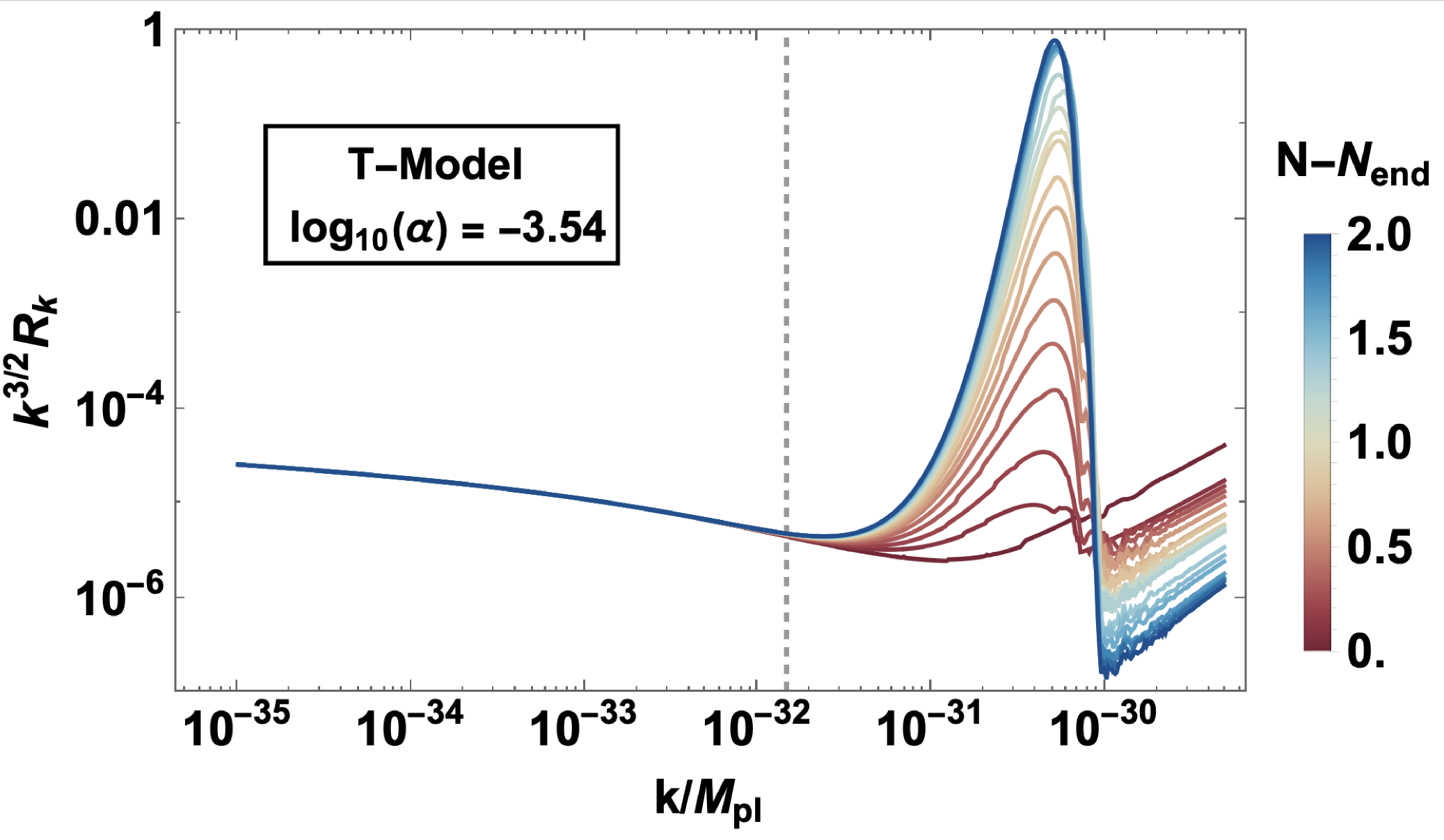}}
    \subfigure[]{\includegraphics[width=0.95\linewidth]{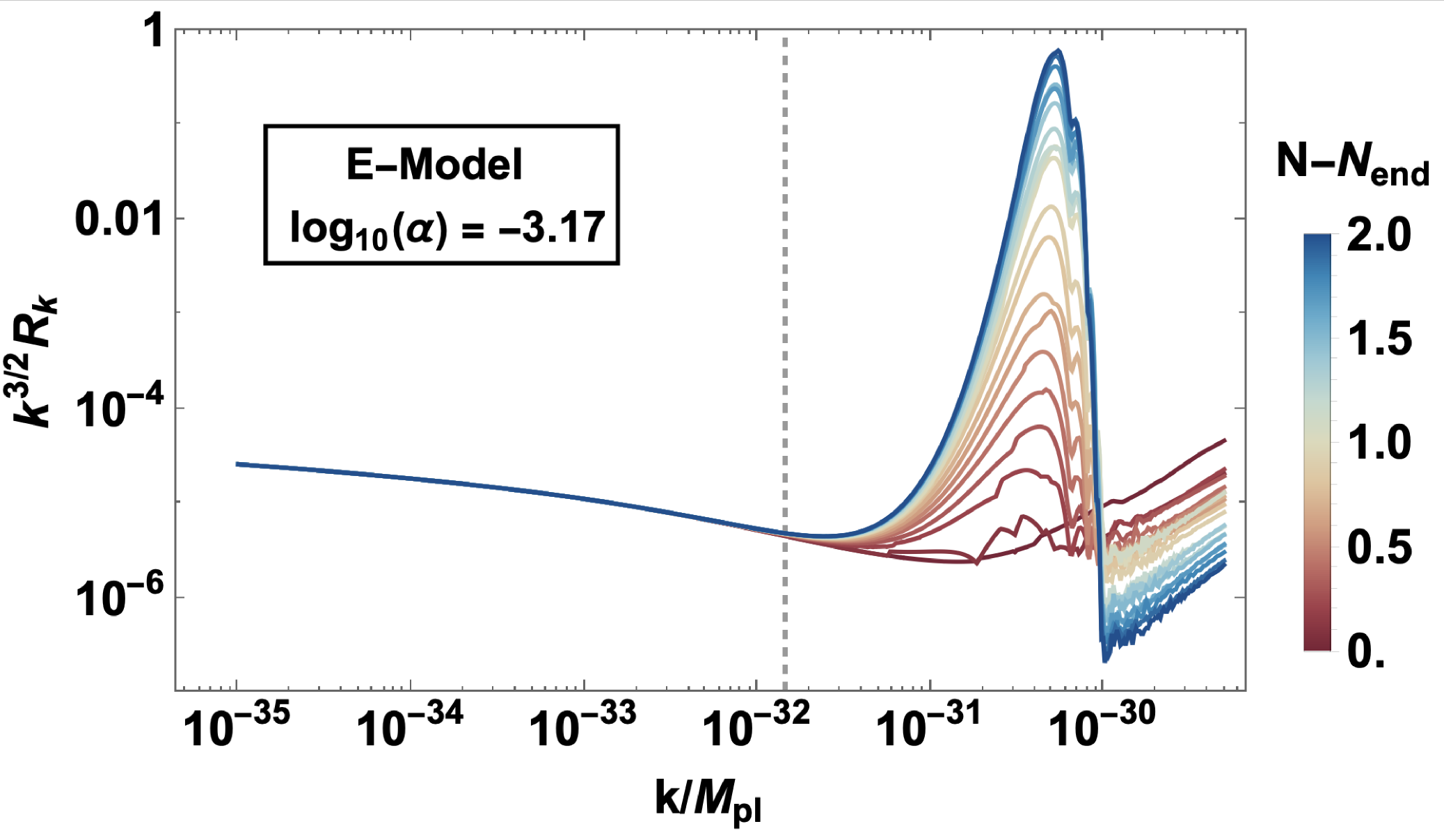}}
    \caption{Curvature perturbation $\mathcal{R}_{\bm k}$ as a function of the comoving wavenumber $k$ for (a) T-model and (b) E-model. The evaluation is performed from the end of inflation to 2 e-folds after in steps of 0.1 e-folds. The vertical dashed lines mark the scale $k_{\text{end}}$.}
    \label{fig:huge-amplification}
\end{figure}

The cosmological evolution of a scalar field minimally coupled to gravity is governed by the action
\begin{equation}\label{eq:lagrangian}
  S= \int d^4x\sqrt{-g}\Bigg[\frac{\Mpl^2}{2}R-\frac12g^{\mu\nu}\partial_{\mu}\phi\partial_{\nu}\phi-V_{T,E}(\phi)\Bigg]
\end{equation}
where $g_{\mu\nu}$ is the space-time metric and $R$ is the Ricci scalar. To study the scalar perturbations, we introduce a fluctuation in the scalar field as $\phi\rightarrow\phi+\delta\phi$, which sources scalar perturbations $\Phi$ and $\Psi$ in the metric (Bardeen potentials), given by the following perturbed FLRW line element
\begin{equation}
    \dd s^2=a^2[-(1+2\Phi)\dd \eta^2+(1-2\Psi)\delta_{ij}\dd x^i\dd x^j].
\end{equation}
Here $\delta_{ij}$ is the Kronecker delta, and we have chosen to work in the Newtonian gauge. Since the scalar field $\phi$ does not introduce any anisotropic stress, the two Bardeen potentials $\Phi$ and $\Psi$ are equal, $\Phi=\Psi$.
Considering that the matter content is described by the scalar field $\phi$, then the solution to the Einstein equations in this perturbed FLRW metric leads to the Mukhanov-Sasaki (MS) equation in Fourier space
\begin{equation}\label{eq:MS-tau}
    v''_{\bm k}(\eta)+\left[k^2-\frac{\left(a\sqrt{\epsilon_H}\right)''}{a\sqrt{\epsilon_H}}\right]v_{\bm k}(\eta)=0,
\end{equation}
where a tilde means derivation with respect to conformal time $\eta$, related to the cosmic time as $\dd\eta=\dd t/a$. Also,
\begin{equation}
    \epsilon_H= -\frac{\dot{H}}{H^2}= \frac12\left(\frac{\phi^\prime}{\mathcal H}\right)^2, \qquad \mathcal H=\frac{a'}{a},
    \label{slow-roll}
\end{equation}
are the first slow-roll parameter and the conformal Hubble rate, respectively, and 
\begin{equation}\label{eq:MSvariable}
v_{\bm k}=a\left[\delta\phi_{\bm k}+\frac{\phi'\Phi_{\bm k}}{\mathcal{H}}\right]
\end{equation}
is the so-called MS variable, a combination of the field and metric perturbations, $\delta\phi_{\bm k}$ and $\Phi_{\bm k}$, respectively. See \cite{Mukhanov:1990me,Baumann:2009ds} for details. As shown in \cite{del-Corral:2024vcm}, during a self-resonant phase, one can perturbatively solve the evolution equation of the inflation field as a series expansion of a small parameter $\beta$ as follows 
\begin{equation}
\phi(\tau)=\sum_{n=1}^{\infty}\beta^n\phi_n(\tau),
\end{equation}
where $\tau=M\sqrt{1-\beta^2}\,t$. Then, the MS equation \eqref{eq:MS-tau} can be transformed into the following Hill equation 
\begin{equation}\label{eq:hill}
    \frac{\dd^2 v_{\bm k}}{\dd\tau^2}+\left[A_k+\sum_{n=1}^{\infty}\Big(q_n(\tau)\cos(n\tau)+p_n(\tau)\sin(n\tau)\Big)\right]v_{\bm k}=0,
\end{equation}
where the functions $A_k$, $q_n$, and $p_n$ depend upon the parameters of the model, and the potential has been expanded in powers of $\phi$. Using the Floquet theorem, the MS variable in \eqref{eq:hill} is found to evolve as $v_{\bm k}\sim\exp{[\int\mu_k\dd\tau]}$, where $\mu_k$ are the so-called Floquet exponents, which in turn depend on the functions $A_k$, $q_n$ and $p_n$. For $\Re(\mu_k)>0$, we have exponential amplification, and thus the mode is said to be unstable, which translates into a strong amplification for a particular range of modes. To show this, we relate the MS variable to the curvature perturbation $\mathcal R_{\bm k}$ as follows
\begin{equation}
\mathcal{R}_{\bm k}=\frac{v_{\bm k}}{\sqrt{2\epsilon_H}a},
\end{equation}
whose time-evolution is shown in {Fig.~\ref{fig:huge-amplification}.} {We observe that}, as $\alpha$ decreases, the curvature perturbation $\mathcal{R}_{\bm k}$ {rapidly} amplifies during the first e-folds after the end of inflation, caused by the self-resonance effect. {This is essentially what characterizes the robustness of our result. Even a short duration of preheating of 1-2 e-folds will produce this amplification for the perturbations. Fig.~\ref{fig:huge-amplification} shows the curvature perturbations for a T- and an E-model with the values of $\alpha$ corresponding to the lower bounds derived in this work. The evaluation is performed from the end of inflation to 2 e-folds after, in steps of 0.1 e-folds. One can observe that the amplification is significant even for a relatively short duration of preheating, which confirms our statement. If one extends the evaluation of this plot further than 2 e-folds after the end of inflation, the curvature perturbations would not increase, as the self-resonance effects are only present during the first e-folds after inflation.}

{Using the perturbed Einstein equations in the Newtonian gauge, one can obtain the following set of equations that allows us to obtain the metric and the field perturbations 
\begin{subequations}
\begin{equation}\label{eq:metric-pert}
    \mathcal{R}_{\bm k}=\frac{H^{-1}\dot{\Phi}_{\bm k}+\Phi_{\bm k}}{\epsilon_H}+\Phi_{\bm k},
\end{equation}
\begin{equation}\label{eq:field-pert}
    \delta\phi_{\bm k}=\sqrt{2\epsilon_H}\left(\mathcal{R}_{\bm k}-\Phi_{\bm k}\right).
\end{equation}
\end{subequations}
These perturbations also carry the self-resonance effects from the curvature perturbation and, as we show below, are essential for the computation of the SIGWs.}

This amplification is crucial for scalar-induced gravitational waves in the second order. Note that we study all the modes within the instability band (See Fig.~\ref{fig:scheme}), which constitute those that exit the horizon during inflation as well as those that never exit the horizon but fall into the instability band with wavelengths greater than the Jeans length $R_J\simeq\sqrt{2\langle c_s\rangle^2/3H^2}$ \cite{del-Corral:2023apl,draft-alpha-PBH} where \begin{equation}\label{eq:speed-of-sound-numerical}
    \langle c_s^2(k)\rangle=\frac{\langle\delta p_{\bm k}\rangle}{\langle\delta\rho_{\bm k}\rangle}\simeq\frac{\langle\frac{k^2}{a^2}\phi-V'(\phi)+V''(\phi)\phi\rangle}{\langle\frac{k^2}{a^2}\phi+3V'(\phi)+V''(\phi)\phi\rangle},
\end{equation}
where $\langle\dots\rangle$ means averaging over a time period larger than the oscillation period but smaller than the Hubble time \cite{Cembranos:2015oya}. The potential $V(\phi)$ in \eqref{eq:speed-of-sound-numerical} is the inflaton potential. For the E-model and T-model alpha-attractor potentials \eqref{eq:T-model}, the analytical estimate for the average sound speed is 
\begin{equation}\label{eq:speed-of-sound-analytical}
    \langle c_s^2(k)\rangle\simeq\frac{k^2+2\lambda a^2\langle\phi^2\rangle}{k^2+4M^2a^2+6\lambda a^2\langle\phi^2\rangle}.
\end{equation}
where $\lambda =-\frac{2M^2}{9\alpha}<0$ for the T-model and $\lambda = \frac{7M^2}{9\alpha}>0$ for the E-model. 
It was shown in \cite{del-Corral:2023apl,draft-alpha-PBH} that both the modes that exit and those that never exit the horizon during inflation could end up inside the instability band with identical behavior. 
However, the modes that never exit the horizon during inflation and enter the instability band get amplified largely and quickly. As we can see in Fig.~\ref{fig:huge-amplification}, the amplification is significant (a few orders of magnitude higher) for the modes with $k>k_{\rm end}$ compared to those with $k<k_{\rm end}$, where $k_{\rm end}$ is the wave number of the mode that exits the horizon at the end of inflation, {marked by the vertical dashed lines}. This amplification of curvature perturbations is essential for computing the SIGWs, which we present in the next section.


\section{Tensor perturbations}\label{sec:tensor-perturbations}

After computing the scalar perturbations during preheating, we aim to analyze the tensor perturbations to determine whether the amplification of scalar perturbations leaves an imprint on the power spectrum of SIGWs. We first use perturbation theory in Sec.~\ref{sec:energy-density} to estimate the energy density of gravitational waves and then, in Sec.~\ref{sec:numerical-strategy}, we explain the numerical strategy followed to achieve that goal.


\subsection{Energy density of gravitational waves}\label{sec:energy-density}

{Going to the 3rd order perturbation of the action \eqref{eq:lagrangian} in scalar ($\Phi,\,\delta\phi$) and tensor  ($h_{ij}^{\lambda}$) fluctuations, we obtain the second-order \footnote{At the level of first order perturbed Einstein equation (or at the level of second order action) scalar and tensor modes do not couple, whereas the perturbed Einstein equations at the second-order (i.e., at the level of 3rd order perturbed action) we can have scalar modes acting as additional source to the generation of gravitational waves on top of the background as \eqref{eq:SOEOM} illustrates.} equations of motion for the tensor modes $h_{\textbf{k}}$ in Fourier space as} \cite{Domenech:2021,Baumann:2007,Ananda:2006,Assadullahi:2009}
\begin{equation}\label{eq:SOEOM}
    \ddot h_{\bm k}^{(\lambda)}(t)+3H\dot h_{\bm k}^{(\lambda)}(t)+\frac{k^2}{a^2}h_{\bm k}^{(\lambda)}(t)= \frac{\mathcal{S}(\bm k,t)}{a^2},
\end{equation}
where $\lambda=+,\times$ denotes the two polarization states, a dot means derivative with respect to cosmic time, and $\mathcal{S}(\bm k,t)$ is the source term, given by
\begin{equation}
\begin{split}\label{eq:source-term}
    \mathcal{S}(\bm k,t)
    &=\int\frac{\dd^3\tilde{k}}{(2\pi)^{3/2}}\tilde k^2(1-\mu^2)\\&\times\left(4\Phi_{\tilde{\bm{k}}}\Phi_{\bm{k}-\tilde{\bm{k}}}+2\delta\phi_{\tilde{\bm{k}}}\delta\phi_{\bm{k}-\tilde{\bm{k}}}\right).
\end{split}
\end{equation}
The above source term contribution originates from the scalar-scalar-tensor vertex interaction term in the third-order perturbed expansion of the action \eqref{eq:lagrangian}{, which is 
\begin{equation}
\begin{aligned}
\delta^{(3)}S  = \int d^3x\, dt \Bigg\{
- a\, h^{ij}\,\partial_i\Phi\,\partial_j\Phi
+ \frac{a}{2}\, h^{ij}\,\partial_i\delta\varphi\,\partial_j\delta\varphi
\Bigg\}.
\end{aligned}
\label{3rdordac}
\end{equation}}
Here, we neglect 4th and higher-order contributions on the expectation that any further scalar self-interactions are negligible. {It is important to emphasize that the source term in Eq.~\eqref{eq:source-term} is expressed in terms of the gauge-invariant variables $\Phi_{\bm{k}}$ and $\delta\phi_{\bm{k}}$ \cite{Mukhanov:1990me}, as a result, no gauge ambiguities are expected to arise.} As we can see from Fig.~\ref{fig:huge-amplification}, the curvature perturbation, which is a combination of metric and scalar field fluctuation, remains smaller than unity even in the regime of self-resonance amplification during the preheating phase. If we consider the scales of the full non-linear regime, then the effects of non-Gaussianities are important \cite{Papanikolaou:2024kjb,He:2024luf}, which we defer for future investigations. Since the source term \eqref{eq:source-term} on the right-hand side of \eqref{eq:SOEOM} reflects the fact that these GWs are no longer free-propagating waves but rather a metric distortion arising from terms quadratic in the scalar perturbations \cite{Assadullahi:2009}, they are called SIGWs. Moreover, the integral over $\tilde{k}$ reflects that the computation of SIGWs involves a convolution of different modes, meaning the contribution from any individual mode is diluted and mixed with contributions from other modes. Consequently, the amplification of the gravitational waves arises from the collective contribution of all modes within the Hubble horizon. In our case, the significant contribution to the amplification comes from the modes $k>k_{\rm end}$ (See Fig.~\ref{fig:huge-amplification}). In App.~\ref{sec:EOM-SIGWs} we show the derivation of the source term $\mathcal{S}(\bm k,t)$ from the perturbed Einstein equations and how to express it in terms of the metric scalar perturbations, as shown in eqn.~\eqref{eq:source-term}.

We can compute the energy density of GWs as the 00 component of the pseudo-energy momentum tensor of GWs \cite{Domenech:2021}, that is
\begin{equation}\label{eq:energy-density}
\begin{split}
    \rho_{\text{GW}}&=\Mpl^2\int\dd\ln{k}\frac{k^3}{16\pi^2}\\&\times\Bigg\{\sum_\lambda\langle\dot{h}_{\bm{k}}^{(\lambda)}\dot{h}_{-\bm{k}}^{(\lambda)}\rangle^!+\frac{k^2}{a^2}\langle h_{\bm{k}}^{(\lambda)}h_{-\bm{k}}^{(\lambda)}\rangle^!\Bigg\},
\end{split}
\end{equation}
where the notation ``!'' indicates that the Dirac delta has been factored out and thus $\langle h_{\bm{k}}^{(\lambda)}h_{-\bm{k}}^{(\lambda)}\rangle^!$ is directly related to the power spectrum $\mathcal P_h(k,t)$, defined as
\begin{equation}\label{eq:spectrum-definition}
    \langle h_{\bm k}^{(\lambda)}(t)h_{\bm k'}^{(\lambda)}(t)\rangle=\frac12\frac{2\pi^2}{k^3}\delta^3(\bm k+\bm k')\mathcal P_h (k,t),
\end{equation}
where the $1/2$ factor comes from the fact that $\mathcal P_h(k,t)$ includes contributions from both polarizations. For our case, this power spectrum can be computed as follows
{\begin{equation}\label{eq:power-SIGWs}
\begin{split}
    \mathcal{P}_h^{\text{SI}}(k,t)&=\frac{16g^2(k,t)}{k}\int_0^{\infty}\dd \tilde{k}\int_{-1}^1\dd\mu\,\frac{\tilde{k}^3\,(1-\mu^2)^2}{|k-\tilde k|^3}\\&\times\bigg[\mathcal P_{\Phi}(\tilde k)\mathcal P_{\Phi}(|\bm k-\tilde{\bm k}|)+\frac12\mathcal P_{\delta\phi}(\tilde k)\mathcal P_{\delta\phi}(|\bm k-\tilde{\bm k}|)\bigg],
\end{split}
\end{equation}}
where SI stands for scalar-induced, $\mu$ is the cosine of the angle between ingoing and outgoing scales, and $\mathcal P_{\Phi}(k)$ { and $\mathcal P_{\Phi}(k)$ are the power spectra of the scalar $\Phi_{\bm k}$ and field $\delta\phi_{\bm k}$ perturbations. They are} defined similarly to \eqref{eq:spectrum-definition}, but a Heaviside function is introduced in this case to avoid the contribution to the integral coming from the non-linear regime \cite{Assadullahi:2009}, that is
\begin{equation}
    \langle \Phi_{\bm k}(t)\Phi_{\bm k'}(t)\rangle=\frac{2\pi^2}{k^3}\delta^3(\bm k+\bm k')\mathcal P_{\Phi} (k,t) \Theta(k_U-k),
    \label{eq:phi-PS}
\end{equation}
where the function $g(k,t)$ is defined in eqn.~\eqref{eq:particular-solution}, {and the same occurs for the field perturbations.} App.~\ref{sec:PS-SIGWs} shows the details about the derivation of \eqref{eq:power-SIGWs} following the standard derivations of GWs production during matter-dominated scenarios as well as some analytical estimates for the spectral energy density of GWs. The latter, labeled as $\Omega_{\text{GW}}$, is often used instead of $\rho_{\text{GW}}$ to compare theoretical predictions with current constraints and future observations. It represents the energy density per logarithm of $k$ over the critical density and is given by
\begin{equation}\label{eq:energy-density-GWs}
    \Omega_{\text{GW}}(k,t)=\frac1{3\Mpl^2H^2}\frac{\dd \rho_{\text{GW}}}{\dd\ln{k}}.
\end{equation}
After the preheating stage, where the radiation-dominated (reheating) period begins, the GWs behave as freely propagating waves. We assume that this transition occurs suddenly. This implies that the following approximation holds from the end of preheating onwards
\begin{equation}\label{eq:app-free}
    \dot{h}_{\bm{k}}^{(\lambda)}\simeq i\frac{k}{a}h_{\bm{k}}^{(\lambda)}.
\end{equation}
Thus using eqns.~\eqref{eq:energy-density}, \eqref{eq:energy-density-GWs} and \eqref{eq:app-free}, the spectral energy density of SIGWs at the end of preheating is given by
\begin{equation}
    \Omega_{\text{GW}}^{\text{SI}}(k,t_{\text{rh}})=\frac{\mathcal{P}_{h}^{\text{SI}}(k,t_{\text{rh}})}{12}\left(\frac{k}{k_L}\right)^2,
\end{equation}
where we have used $a(t_{\text{rh}})H(t_{\text{rh}})=k_L$. After the end of preheating, the fractional energy density redshifts at sub-Hubble scales as any non-interacting relativistic particles. Therefore, the present energy density of gravitational waves is approximated by \cite{Assadullahi:2009}
\begin{widetext}
{\begin{equation}\label{eq:energy-density-final}
\begin{split}
    \Omega_{\text{GW}}^{\text{SI}}(k,t_0)&\simeq\frac{4\,\Omega_\gamma^0}{3}\left(\frac{k}{k_{L}}\right)^2\frac{g^2(k,t_{\text{rh}})}{k}\int_0^{\infty}\dd \tilde{k}\int_{-1}^1\dd\mu\,\frac{\tilde{k}^3(1-\mu^2)^2}{|k-\tilde k|^3}\bigg[\mathcal P_{\Phi}(\tilde k)\mathcal P_{\Phi}(|k-\tilde k|)+\frac12\mathcal P_{\delta\phi_{\bm k}}(\tilde k)\mathcal P_{\delta\phi_{\bm k}}(|k-\tilde k|)\bigg]\\
    &\simeq \Omega_{\text{GW},\Phi}^{\text{SI}}(k,t_\text{rh})+\Omega_{\text{GW},\delta\phi}^{\text{SI}}(k,t_\text{rh}) .
\end{split}
\end{equation}}
\end{widetext}
where $\Omega_{\gamma}^0\simeq1.2\times10^{-5}$ is the present energy density of photons, $t_0$ represents the present epoch, and we have substituted in the last line the power spectrum from eqn.~\eqref{eq:power-SIGWs}, see \cite{Assadullahi:2009} for details. This quantity is usually expressed as a function of the GW frequency $f$ rather than the wave number $k$. The relationship between them is given by \cite{Frosina:2023nxu}
\begin{equation}
    f=1.55\times10^{-15}\left(\frac{k}{\Mpc^{-1}}\right)\Hz.
\end{equation}

As previously mentioned, we are also interested in the BGWs, whose evolution equation is also \eqref{eq:SOEOM} but with $\mathcal{S}_{\bm k}=0$, indicating that BGWs behave as free waves. Inflation predicts an almost scale-invariant power spectrum of BGWs, given by
\begin{equation}\label{eq:background-tensor-spectrum}
    \mathcal P^{\text{inf}}_h(k)=\mathcal{A}_h\left(\frac{k}{k_*}\right)^{n_t},
\end{equation}
where $k_*=0.05\Mpc^{-1}$ is the pivot scale, $\mathcal{A}_h$ is the tensor amplitude, and $n_t$ the tensor spectral index. These parameters are related to the tensor-to-scalar ratio \eqref{vpred} as $\mathcal{A}_h=r\mathcal{A}_{\mathcal{R}}$ and $n_t=-r/8$, where $\mathcal{A}_{\mathcal{R}}=2.2\times10^{-9}$ is the amplitude of the scalar power spectrum at the pivot scale. Let us begin by studying the evolution of eqn.~\eqref{eq:background-tensor-spectrum} after inflation. To analyze this, let us solve \eqref{eq:SOEOM} (with $\mathcal{S}_{\bm k}=0$) under different expansion histories of the universe. Defining $h_{\bm k}=\tilde h_{\bm k}/a$ (omitting the polarization index $\lambda$) and using conformal time $\eta$, we can express \eqref{eq:SOEOM} as
\begin{equation}
    \tilde h_{\bm k}''+\left(k^2-\frac{\nu^2-\frac14}{\eta^2}\right)\tilde h_{\bm k}=0,
\end{equation}
where $\nu=\frac{3(1-w)}{2(1+3w)}$, and we have assumed that $a\sim\eta^{\frac{2}{1+3w}}$, with $w$ being the equation of state parameter of the universe. The solution to this equation is given in terms of Bessel functions:
\begin{equation}
    h_{\bm k}=\sqrt{\eta}\left[A_kJ_{\nu}(k\eta)+B_kY_{\nu}(k\eta)\right].
\end{equation}
During preheating (reheating), we have $\nu \simeq 3/2$ ($\nu = 1/2$). Thus, for sub-horizon modes, and in both cases, $J_{\nu}(k\eta) \sim Y_{\nu}(k\eta) \sim 1/\sqrt{k\eta}$, implying $h_{\bm{k}} \sim 1/a$. This allows us to define the transfer function for the background tensor perturbations as:
\begin{equation}
    \mathcal{T}^{\,2}_h(k,t) = \begin{cases}
        1 & t<t_k \\
        \frac{1}{2} \left(\frac{a_k}{a}\right)^2 & t>t_k
    \end{cases}
\end{equation}
where $a_k$ is the scale factor at the time the mode $k$ enters the horizon, defined as $t_k$. The factor $1/2$ accounts for averaging over harmonic oscillations of modes deep inside the horizon~\cite{Figueroa:2019paj}. The background tensor power spectrum evolves as follows after inflation:
\begin{equation}
    \mathcal{P}_h^{\text{B}}(k,t) = \mathcal{P}_h^{\text{inf}}(k) \mathcal{T}_h^2(k,t),
\end{equation}
and the background fractional energy density is computed using \eqref{eq:energy-density-GWs} as
\begin{equation}\label{eq:omega-pre}
    \Omega^{\text{B}}_{\text{GW}}(k,t) = \frac{k^2}{12a^2H^2} \mathcal{P}_h^{B}(k,t) = \frac{\mathcal{P}_h^{\text{inf}}(k)}{24} \frac{a_k^4 H_k^2}{a^4 H^2}.
\end{equation}
If the mode enters during preheating, a suppression effect exists due to the $a^4 H^2 \propto a$ term in eqn.~\eqref{eq:omega-pre}. This means that the fractional energy density gets suppressed as $\Omega^{\text{B}}_{\text{GW}} \propto a_k / a$ for the modes entering during preheating \cite{Figueroa:2019paj,Caprini:2018mtu,Sahni:2001qp,Sahni:1990tx,Mishra:2021wkm}, that is, for $k>k_L$. This effect disappears when reheating begins. At this point, and as in the scalar-induced case, the fractional energy density redshifts as any non-relativistic particles. This implies we can use the following parametrization for the fractional energy density of BGWs evaluated at the present time:
\begin{equation}
    \Omega_{\text{GW}}^{\text{B}}(k,t_0) = \Omega_\gamma^0 \frac{r \mathcal{A}_{\mathcal{R}}}{24} \left(\frac{k}{k_*}\right)^{n_t}
    \begin{cases} 
        \frac{a_k}{a_\text{rh}}, & \text{if } k > k_{L} \\ 
        1, & \text{if } k \leq k_{L}
    \end{cases}
    \label{eq:BGW}
\end{equation}
Further suppression effects exist for modes entering during the late matter- and dark-energy-dominated eras. However, the associated wavelengths are much smaller than those relevant during preheating and are, therefore, not considered here. Finally, the total fractional energy density of GWs evaluated at present is the sum of both contributions (i.e., \eqref{eq:BGW} and \eqref{eq:energy-density-final})
\begin{equation}\label{eq:total-omega}
    \Omega_{\text{GW}}^{\text{TOT}}(k,t_0) = \Omega_{\text{GW}}^{\text{B}}(k,t_0) + \Omega_{\text{GW}}^{\text{SI}}(k,t_0),
\end{equation}
where the numerical computation of $\Omega_{\text{GW}}^{\text{SI}}$ is detailed in the next subsection.


\subsection{Numerical strategy}\label{sec:numerical-strategy}

To numerically compute the total fractional energy density of SIGWs, we begin by computing the time evolution of the curvature perturbations $\mathcal{R}_{\bm k}$, following the method outlined in Sec.~\ref{sec:scalar-perturbations}. {Then, using eqns.~\eqref{eq:metric-pert} and \eqref{eq:field-pert} we compute the metric and field perturbations $\Phi_{\bm k}$ and $\delta\phi_{\bm k}$}. The results are stored in $k$-space. Next, the power spectrum of the {metric and field} perturbations,  {$\mathcal P_{\Phi}$ and $\mathcal P_{\delta\phi}$}, {are} computed and evaluated at the end of the short self-resonance stage, as it remains constant for the modes of interest thereafter. 
Since the {scalar} perturbations can only be obtained numerically in discrete steps of $k$, the integral over $\tilde{k}$ in \eqref{eq:energy-density-final} is discretized. Specifically, the $k$-space integral is divided into 100 uniform points per logarithmic interval of $k$. This procedure is repeated for each value of $k$ in $k\in[k_{L},k_{U}]$ and evaluated at the end of the preheating stage, $t_{\text{rh}}$, which we consider to {range from 1 to 5 e-folds after the end of inflation}. 

\begin{figure}
         \centering
         \includegraphics[width=\linewidth]{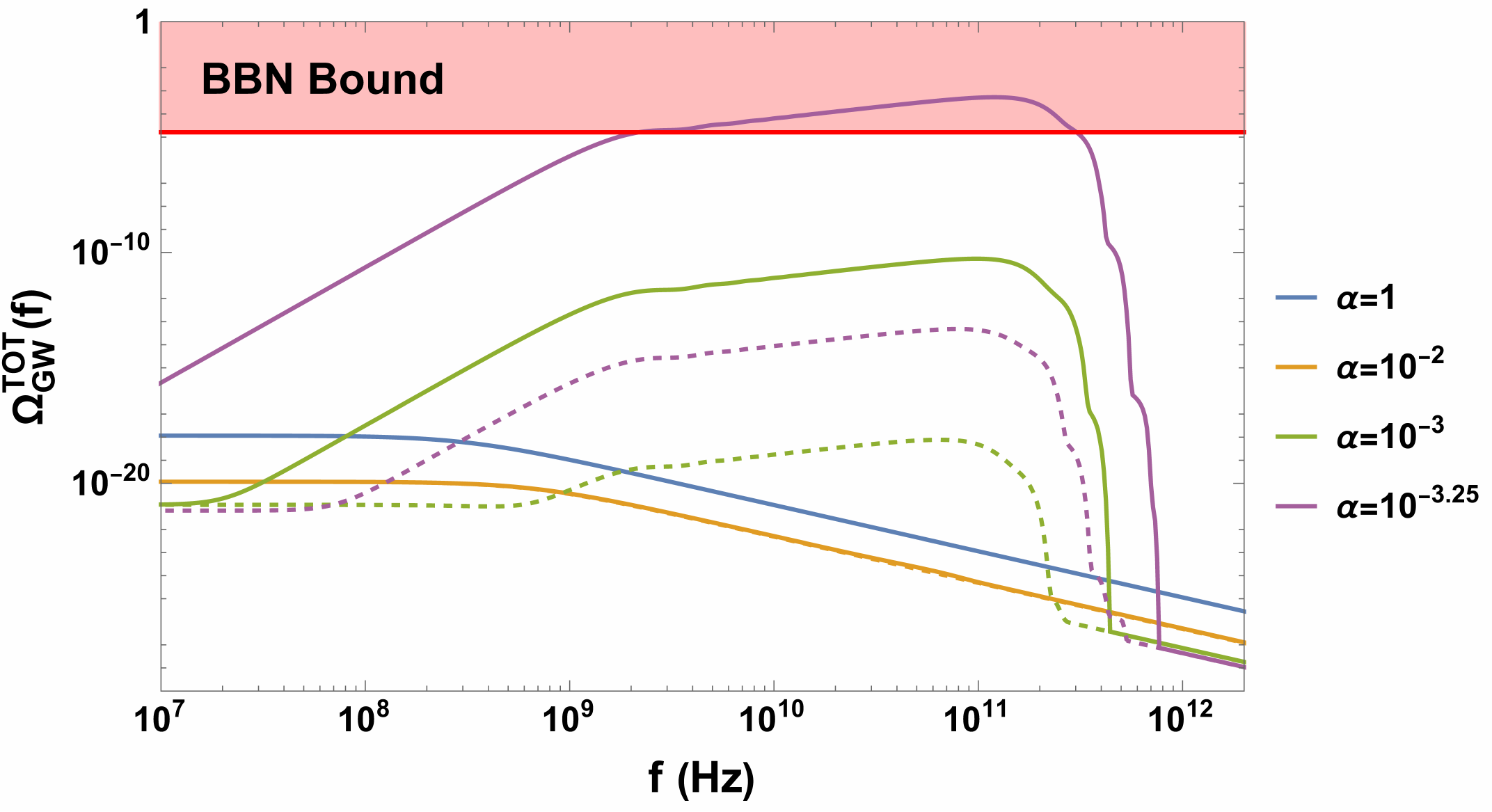}
        \caption{Total fractional energy density of GWs, eqn.~\eqref{eq:total-omega}, evaluated today for a T-model (dashed lines) and an E-model (solid lines), with different values of $\alpha$ and as a function of the frequency of each mode, expressed in Hz. The red horizontal line represents the BBN bound. Preheating is assumed to last for 5 e-folds.}
        \label{fig:Omega}
\end{figure}

Fig.~\ref{fig:Omega} shows the computation of the total fractional energy density of GWs for the two models under consideration (T-model as dashed lines and E-model as solid lines) and for different values of the parameter $\alpha$. {The length of preheating is chosen as 5 e-folds in this case}. We have also included the Starobinsky model \cite{Starobinsky:1980te}, which corresponds to an E-model with $\alpha=1$, for comparison purposes. An interesting feature of Fig.~\ref{fig:Omega} is that the amplification peak for high-$k$ in the scalar perturbations of Fig.~\ref{fig:huge-amplification} is also present in the energy density of GWs for small $\alpha$, since $\Omega_{\text{GW}}^{\text{TOT}}\sim\Omega_{\text{GW}}^{\text{SI}}$. However, for higher values of $\alpha$, such as the Starobinsky case, the amplification of scalar perturbations is negligible and thus $\Omega_{\text{GW}}^{\text{TOT}}\sim\Omega_{\text{GW}}^{\text{B}}$. Nonetheless, due to the convolution mentioned earlier, this contribution from the peak gets diluted among the rest of the modes inside the horizon, and thus the peak in $\Omega_{\text{GW}}^{\text{TOT}}$ is not as sharp as it is in {$\mathcal{P}_{\Phi}$ and $\mathcal{P}_{\delta\phi}$}, which makes it difficult to track back the initial feature in the power spectrum causing the amplification. In App.~\ref{sec:PS-SIGWs} we show some analytical approximations that explain the shape of $\Omega_{\text{GW}}^{\text{SI}}$. We also remark here that, as anticipated, decreasing $\alpha$ leads to a growth of the metric perturbations, which enhances gravitational wave production in the very high-frequency (VHF) band. Although there is no current GW detector for the VHF band, we can use the BBN bound to restrict the model's parameters. This proceeds as follows. The GWs at the onset of the BBN contribute to the radiation density and can potentially change the expansion rate of the universe, and thus the abundance of the light elements. Measurements of this abundance lead to an equivalent number of additional relativistic species of 1.4 neutrino degrees of freedom, which translates into the following upper bound for the total fractional energy density of GWs \cite{Smith:2006nka,Maggiore:1999vm}
\begin{equation}\label{eq:BBN}
    \mathcal I_{\Omega}=\int_{0}^{\infty}\Omega_{\text{GW}}^{\text{TOT}}(k)\,\dd\ln(k)\leq 1.6\times10^{-5}=\mathcal I_{\text{BBN}}.
\end{equation}

\begin{figure}
    \centering
    \subfigure[]{\includegraphics[width=\linewidth]{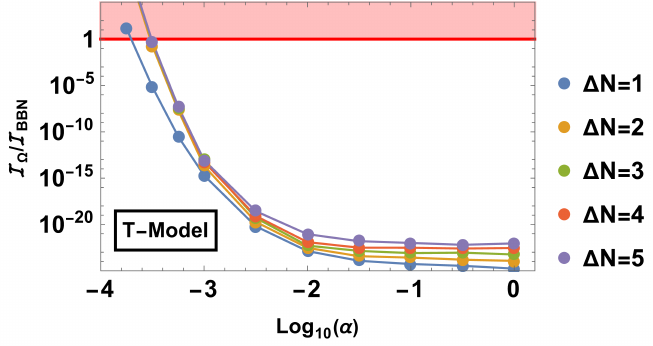}}
    \subfigure[]{\includegraphics[width=\linewidth]{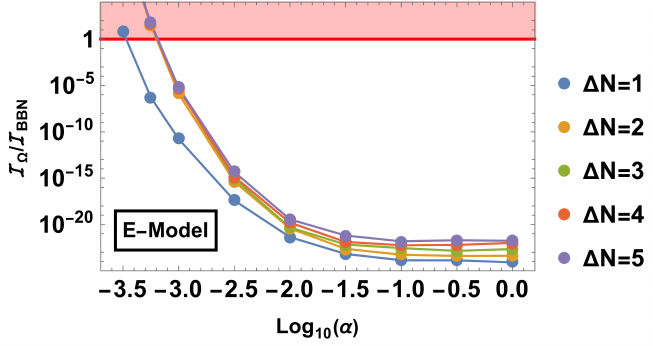}}
    \caption{Integrated fractional energy density, eqn.~\eqref{eq:energy-density-final}, normalized to $\mathcal{I}_{\text{BBN}}$ as a function of $\alpha$ {and the duration of preheating ($\Delta N$)} for the T- and E-models. {One finds that extending preheating beyond 2 e-folds does not affect the moment when the BBN bound is reached. The reason is that after two e-folds all relevant (amplified) perturbations have entered the horizon, making their contribution to the integral \eqref{eq:BBN} negligible.}}
    \label{fig:ints}
\end{figure}
This bound is represented as the horizontal red band in Fig.~\ref{fig:Omega}, which is reached for the E-model with $\alpha=10^{-3.25}$. It is important to remark that the bound is on the integral over the whole spectrum of modes, which implies that $\Omega_{\text{GW}}^{\text{TOT}}$ could present a peak above $\mathcal I_{\text{BBN}}$, provided it is narrow so that $\mathcal I_{\Omega}<\mathcal I_{\text{BBN}}$. Thus, Fig.~\ref{fig:Omega}, is not good enough to establish a lower bound on $\alpha$. Instead, Fig.~\ref{fig:ints} shows the result of the numerical integration of the fractional energy density \eqref{eq:energy-density-final} as a function of $\alpha$ {and the duration of preheating ($\Delta N$)} for the T- and E-models, normalized to $\mathcal{I}_{\text{BBN}}$. Here, we clearly see a limiting value of $\alpha$, {that we call $\alpha_{\text{BBN}}$,} for which the BBN bound (red shaded area) is reached {for each $\Delta N$}.{ The results can be seen in Table~\ref{tab:bounds}.} Since the amplification occurs rapidly ($\mathcal{O}(1)$ e-fold), this essentially restricts the duration of preheating. Also, we can use \eqref{vpred} to put a lower bound on the tensor-to-scalar ratio {, which can be seen in Table~\ref{tab:bounds} as $r_{\text{BBN}}$.} {In general, one can observe that $\alpha_{\text{BBN}}$ is higher for the E-model compared to the T-model for the same $\Delta N$. This} rests on the fact that the E-model potential is asymmetric and thus self-resonance is stronger in this case, for the same value of $\alpha$, due to the drift terms in the potential expansion \cite{del-Corral:2024vcm}. This lower bounds, although far from the sensitivity of the future satellite observer LiteBIRD ($ r\sim1.0\times10^{-3}$) \cite{LiteBIRD:2022nob,LiteBIRD:2023aov}, agrees with the bound using BICEP2 and Planck data \cite{Tristram:2021tvh}. {This bound can be compared with the results of \cite{Easther:2021rdg} where the authors have obtained a bound on the tensor-to-scalar ratio, based on the latest constraints on the running of the scalar spectral index. Although no precise bound for $r$ is given, it is shown that in three-parameter non-trivial slow-roll inflationary models, if $r\lesssim 10^{-4}$, then the running of the spectral index must satisfy $\alpha_s<-10^{-3}$, placing it near the threshold of detectability by upcoming experiments. It is crucial to note that the bound in \cite{Easther:2021rdg} corresponds to non-trivial single-field models that introduce a large running of spectral index due to a specific potential shape (that has a large non-trivial value for the third derivative of the inflaton potential). In contrast, our investigation is about the simplest single field models of universality class \cite{Roest:2013fha}.}

\renewcommand{\arraystretch}{1.5}
\begin{table}[]
\centering
\begin{tabular}{|c|cl|cl|}
\hline
  & \multicolumn{2}{c|}{\textbf{T-Model}}                & \multicolumn{2}{c|}{\textbf{E-Model}}                \\ \hline
\textbf{$\Delta N$} &
  \multicolumn{1}{c|}{\textbf{$\log(\alpha_{\text{BBN}})$}} &
  \textbf{$r_\text{BBN}$} &
  \multicolumn{1}{c|}{\textbf{$\log(\alpha_{\text{BBN}})$}} &
  \textbf{$r_\text{BBN}$} \\ \hline
1 & \multicolumn{1}{c|}{-3.7071} & 6.5430$\times10^{-7}$ & \multicolumn{1}{c|}{-3.4763} & 1.1132$\times10^{-6}$ \\ \hline
2 & \multicolumn{1}{c|}{-3.5236} & 9.9834$\times10^{-7}$ & \multicolumn{1}{c|}{-3.2027} & 1.0313$\times10^{-6}$ \\ \hline
3 & \multicolumn{1}{c|}{-3.5114} & 1.0268$\times10^{-6}$ & \multicolumn{1}{c|}{-3.1969} & 2.1182$\times10^{-6}$ \\ \hline
4 & \multicolumn{1}{c|}{-3.5094} & 1.0315$\times10^{-6}$ & \multicolumn{1}{c|}{-3.1967} & 2.1192$\times10^{-6}$ \\ \hline
5 & \multicolumn{1}{c|}{-3.5095} & 1.0313$\times10^{-6}$ & \multicolumn{1}{c|}{-3.1924} & 2.1403$\times10^{-6}$ \\ \hline
\end{tabular}
\caption{{Lower bound on $\alpha$ and the corresponding tensor-to-scalar ratio $r$ for each duration of preheating $\Delta N$ from 1 to 5 e-folds.}}
\label{tab:bounds}
\end{table}

In  \cite{Inomata:2019zqy,Inomata:2019ivs}, it is pointed out that the spectrum of SIGWs corresponding to the modes $k<k_{\rm end}$  could be affected by the later transition from the preheating (inflaton matter domination) to the radiation era. In the case of a gradual transition from preheating to radiation, Ref.~\cite{Inomata:2019zqy} indicates that the SIGWs enhancement gets suppressed because of the subsequent decay of the gravitational potential $\Phi_k$. 
However, in the case of a sudden transition \cite{Inomata:2019ivs}, the same conclusion does not apply; in fact, there is an additional enhancement in SIGW. By comparison to \cite{Inomata:2019zqy,Inomata:2019ivs}, in this paper, we also consider in addition the self-resonance effects in the $k> k_{\rm end}$  region of the spectrum (i.e., the modes that never exit the horizon during inflation but fall into the instability band). Notably  the enhanced contribution to SIGW $\Omega_{\text{GW}}^{\text{TOT}}$ is dominated by the modes $k\gg k_{\rm end}$ as we can notice from Fig.~\ref{fig:huge-amplification} and Fig.~\ref{fig:Omega}. 
A recent study of this regime \cite{Chakraborty:2025zgx} shows the gravitational particle production contribution from the small wavelength modes $k \gg k_{\rm end}$ is negligible because the vacuum associated with these modes during inflation is closer to the Minkowski vacuum rather than the de Sitter vacuum. This indicates that the decay of gravitational potential $\Phi_k$ for these modes is unlikely to happen. However, a precise analytical and numerical study of this scenario is left for future work.


\section{Conclusions}\label{sec:conclusions}

In this work, we explored the production of Scalar-Induced Gravitational Waves (SIGWs) during the preheating phase in the context of $\alpha$-attractor inflationary models. 
These models, characterized by the parameter $\alpha$, deviate from a quadratic potential as $\alpha$ decreases (see Fig.~\ref{fig:potentials}), which triggers a self-resonance effect and rapidly and substantially amplifies the scalar perturbations during preheating \cite{del-Corral:2024vcm,Hertzberg:2014iza,Hertzberg:2014jza}. We particularly considered all the perturbations that fall into the preheating instability band, which include both kinds of modes that exit and never exit the horizon during inflation, i.e., $k<k_{\rm end}$ and $k>k_{\rm end}$. In particular, we found that the modes $k>k_{\rm end}$ whose wavelength is larger than the Jeans length are significantly amplified, and these modes would not contribute to particle production in the post-preheating phase. This amplification, in turn, directly impacts the production of gravitational waves, since scalar perturbations couple to the tensor perturbations at second order through the source term (See ~\eqref{eq:SOEOM} and \eqref{eq:source-term}). We computed the total fractional energy density of GWs (SIGWs + BGWs) for different values of $\alpha$ {and durations of preheating ($\Delta N$),} and our findings show that, for sufficiently small values of $\alpha$, the integral of the total fractional energy density of GWs reaches the BBN bound \eqref{eq:BBN}, see Figs.~\ref{fig:Omega} and \ref{fig:ints}. In particular, we observe that the growth in $\Omega_{GW}^{\text{TOT}}$ is extremely large (by several orders of magnitude) for small differences in $\alpha$. See for instance Fig.~\ref{fig:Omega} for $\alpha=10^{-2}-10^{-3}$. This is due to the fact that the fractional energy density of SIGWs is, naively speaking, proportional to $\int\mathcal P_\Phi^2(k)\sim\int\mathcal{P}_{\delta\phi}^2(k)$ (See ~\eqref{eq:energy-density-final}), meaning small variations in {the perturbations} induce large changes in $\Omega_{GW}^{\text{SI}}$ so that the secondary effect (SIGWs) dominates over the linear effect (BGWs). In contrast, this amplification mechanism is suppressed for higher values of $\alpha$, including the Starobinsky case, where the effect becomes negligible.

We demonstrated, for the first time, that the lower limit on $\alpha$ imposed by the BBN constraint differs between T-models and E-models when considering the production of SIGWs during preheating instabilities. This approach relies on the assumption that the decay of the inflaton field is not instantaneous, with its coupling to the fields responsible for reheating the universe being sufficiently small to ensure that the preheating phase lasts at least 1 or 2 e-folds before the inflaton decays into ultra-relativistic particles~\cite{Kofman:1994rk,Kofman:1997yn}. We also stress the fact that the transition into the standard radiation-dominated universe after the inflaton's decay into particles is not studied in this paper. However, as pointed out by the end of Section~\ref{sec:numerical-strategy}, it is unlikely that this could affect our results. {In addition, we note that our derived bounds on tensor-to-scalar ratio are obtained at the level of linearised tensor fluctuations that include the effect of scalar-scalar-tensor interactions with metric and scalar field matter fluctuations. However, our results are subject to verification with the complete non-linear analysis, including all metric fluctuations at higher orders, which would require techniques beyond the existing frameworks of lattice simulations \cite{Caravano:2021pgc,Aurrekoetxea:2023jwd}. We defer this investigation for future studies.}

The distinction in the lower bound on $\alpha$ {(for the same $\Delta N$)} arises from the additional amplification present in the E-model compared to the T-model, which can be attributed to the asymmetry of the potential~\cite{del-Corral:2024vcm}. These new bounds, although derived under different assumptions, should be compared with those reported in~\cite{Iacconi:2023mnw} based on the duration of reheating along with the current bounds on the scalar spectral index, where $\log_{10}(\alpha) = -4.2^{+5.4}_{-8.6}$ at $95\%$ C.L, in \cite{Alam:2023kia} based on the overproduction of a light moduli field, where $\alpha\lesssim10^{-8}$, or in \cite{Krajewski:2018moi} based on geometrical destabilization of a spectator field, where $\alpha\lesssim10^{-3}$. Furthermore, they establish a new lower bound on the tensor-to-scalar ratio, as given in  {Table~\ref{tab:bounds}}.

In conclusion, our analysis suggests that the amplification of scalar perturbations during self-resonant preheating offers an effective mechanism for generating GWs, particularly in the VHF band. This remarks the need for increasing the sensitivity of gravitational wave detectors in the VHF range that could potentially test these theoretical predictions \cite{Aggarwal:2020olq,Kuroda:2015owv,Page:2020zbr,Dandoy:2024oqg} and offer new insights into the physics of the early universe. Furthermore, we derive for the first time a concrete lower bound on the tensor-to-scalar ratio for the single-field Starobinsky-like inflationary scenarios, which would be an important input for the future probes of B-mode polarization \cite{Belkner:2023duz,CMB-S4:2016ple}. 

\acknowledgments
{We thank the anonymous referee for careful reading and important suggestions to improve the quality of the manuscript.}
D.C. is grateful for the support of grant UI/BD/151491/2021 from the Portuguese Agency Funda\c{c}\~ao para a Ci\^encia e a Tecnologia. The work of P.G. was partially supported by NSF grant PHY-2412829. P.G. thanks Prof. Teruaki Suyama for his kind hospitality at the Institute of Science Tokyo. This research was funded by Funda\c{c}\~ao para a Ci\^encia e a Tecnologia grant number UIDB/MAT/00212/2020 and COST action 23130. K.S.K. acknowledges the support from the Royal Society Newton International Fellowship. We would like to thank David Wands for the very useful discussions.


\appendix

\section{Source term for SIGWs}\label{sec:EOM-SIGWs}

In this appendix, we closely follow the analysis presented in \cite{Domenech:2021} and \cite{Baumann:2007}. By considering both metric and field fluctuations, we can write the following line element for the perturbed FLRW metric
\begin{equation}\label{eq:line-element-tensor-scalar}
\begin{split}
    \dd s^2&=g_{\mu\nu}\dd x^\mu\dd x^\nu\\&=-(1+2\Phi)\dd t^2+a^2[(1-2\Psi)\delta_{ij}+h_{ij}]\dd x^i\dd x^j,
\end{split}
\end{equation}
where $h_{ij}$ is a gauge-invariant, symmetric, transverse, and traceless perturbation tensor. We are working in the Newtonian gauge {and we impose $\Phi=\Psi$ constraint since there is no anisotropic stress involved for the scalar field matter. Thus, the perturbed Einstein equations are given by \cite{Mukhanov:1992,Bauman:2009}}

{\begin{equation}
 -3H\left(\dot{\Phi} + H\Phi\right) + \frac{\nabla^{2}\Phi}{a^{2}}
= \frac{1}{\Mpl^2}\,\delta\rho \, .
\end{equation}}
where
\begin{equation}
    {\delta T^{0}{}_{0} = -\delta\rho
= -\left( \dot{\bar{\phi}}\,\delta\dot{\phi}
- \Phi\,\dot{\bar{\phi}}^{2}
+ V'(\bar{\phi})\delta\phi \right)}
\end{equation}
and 
{\begin{equation}
\begin{aligned}
\dot{\Phi}+H\Phi= \frac{1}{2\Mpl^2}\,\dot{\bar{\phi}}\,\delta\phi
\end{aligned}
\label{Phidelphi}
\end{equation}
The perturbed scalar field equations of motion are
\begin{equation}
    \delta\phi^{\prime\prime} + 2\mathcal{H}\delta\phi^{\prime} - \nabla^2\delta\phi + a^2 V_{,\phi\phi}\delta\phi - 4\bar{\phi}^\prime\Phi^\prime + 2a^2 V_{,\phi}\Phi = 0
\end{equation}
We can notice clearly from \eqref{Phidelphi} that the metric and scalar fluctuations are always related to each other. In Fig.~\ref{fig:MScomponents}, we can see in detail that, during inflation, the amplitude of metric fluctuations would be a lot smaller than the scalar-field fluctuations, while during preheating instability, both the metric and scalar field fluctuations would get amplified in the same manner due to the resonant instability. } 

The most general form for $h_{ij}$ is given by
\begin{equation}
    h_{ij}(x,t)=\sum_{\lambda=+,\times}h^{(\lambda)}(t) \tau_{ij}^{(\lambda)}(x),
\end{equation}
where $\tau_{ij}^{(\lambda)}$ are polarization tensors, which are symmetric, transverse, and traceless, and the index $\lambda=+,\times$ denotes the two polarization states. Using the metric defined in \eqref{eq:line-element-tensor-scalar} we  can derive the following field equation for $h_{ij}$ from the third order perturbative expansion of the action \eqref{eq:lagrangian}  \cite{Guzzetti:2016,Bauman:2009}
\begin{equation}\label{eq:EOM}
    \ddot{h}_{ij}+3H\dot{h}_{ij}-\frac{\nabla^2h_{ij}}{a^2}=-\frac{4}{a^{2}}\mathcal{T}_{ij}^{lm}\mathcal{S}_{lm}.
\end{equation}
Here $\mathcal{S}_{lm}$ is the source term, which contains the information about scalar perturbations and is given by \cite{Domenech:2021}
\begin{equation}
    \mathcal{S}_{lm}=-8\partial_l(\Phi+\Psi)\partial_m\Phi+4\partial_l\Phi\partial_m\Phi+2\partial_l\delta\phi\partial_m\delta\phi.
\end{equation}
The term $\mathcal{T}_{ij}^{lm}$ is the projection tensor, responsible for extracting the transverse and traceless components of any tensor and the terms involving second-order scalar and tensor perturbations \cite{Ananda:2006}. It can be expressed in terms of the polarization tensors $\tau_{ij}^{(\lambda)}$ as \cite{Baumann:2007}
\begin{equation}\label{eq:projection-tensor}
\begin{split}
    \mathcal{T}_{ij}^{lm}\mathcal{S}_{lm}&=\int\frac{\dd^3\mathbf{k}}{(2\pi)^{3/2}}e^{i\mathbf{k}\cdot\mathbf{x}}\\&\times\left(\sum_{\lambda=+,\times}\tau_{ij}^{(\lambda)}(\mathbf{k})\tau^{(\lambda)lm}(\mathbf{k})\right)\mathcal{S}_{lm}(\mathbf{k}),
\end{split}
\end{equation}
where $\mathcal{S}_{lm}(\bm k)$ is the Fourier transform of $\mathcal{S}_{lm}(\bm x)$. Additionally, the Fourier transform of $h_{ij}$ is expressed as
\begin{equation}\label{eq:fourier-h}
    h_{ij}(x,t)=\int\frac{\dd^3\mathbf k}{(2\pi)^3}\sum_{\lambda=+,\times}h_{\mathbf k}^{(\lambda)}(t)\tau_{ij}^{(\lambda)}(\mathbf k)e^{i\mathbf k \cdot \mathbf x}.
\end{equation}
By applying eqns.~\eqref{eq:projection-tensor} and \eqref{eq:fourier-h}, we can transform the equation of motion \eqref{eq:EOM} into Fourier space, resulting in the following expression \cite{Baumann:2007}
\begin{equation}\label{eq:SOEM-Fourier}
    \ddot h_{\bm k}^{(\lambda)}+3H\dot h_{\bm k}^{(\lambda)}+\frac{k^2}{a^2}h_{\bm k}^{(\lambda)}= -\frac{\tau^{lm}\mathcal{S}_{(\lambda)lm}(\bm k)}{a^2}.
\end{equation}
Next, we compute the Fourier transform of $\mathcal{S}_{lm}(\bm x)$, for which we use the following property of the Fourier transforms, called the Convolution Theorem. Consider two functions, $f$ and $g$. The Fourier transform of their product is given by \cite{Arfken:379118}
\begin{equation}
    \mathcal{F}\{f(\bm x)g(\bm x)\}\equiv\frac{1}{(2\pi)^{3/2}}(f\ast g)(\bm k),
\end{equation}
where $\mathcal{F}$ represents the Fourier transform and $\ast$ the convolution, defined as
\begin{equation}
    (f\ast g)(\bm k)=\int\dd^3\tilde{\bm k}\, f(\tilde{\bm k})\, g(\bm k-\tilde{\bm k}).
\end{equation}
Using this, the source term in \eqref{eq:SOEM-Fourier} is now given by \cite{Baumann:2007,Ananda:2006,Assadullahi:2009}
\begin{equation}
\begin{split}\label{eq:source-term-pre-final}
    \mathcal{S}(\bm k,t)&=-\tau^{lm}\mathcal{S}_{lm}(\bm k,t)\\
    &=\int\frac{\dd^3\tilde{k}}{(2\pi)^{3/2}}\tau^{lm}\tilde k_l\tilde k_m\left(4\Phi_{\tilde{\bm{k}}}\Phi_{\bm{k}-\tilde{\bm{k}}}+2\delta\phi_{\tilde{\bm{k}}}\delta\phi_{\bm{k}-\tilde{\bm{k}}}\right),
\end{split}
\end{equation}
 The meaning of \eqref{eq:source-term-pre-final} is that, since the source term consists of products of perturbations that interact in real space and thus their Fourier coefficients mix, the resulting mode of the product is influenced by all possible pairs of modes of the original perturbations, which translates into a convolution in $\bm k$. Using the properties of the projection tensors, we have that \cite{Baumann:2007}
\begin{equation}
    \tau^{(\lambda)lm}\tilde{k}_l\tilde{k}_m=\tilde{k}^2(1-\mu^2),
\end{equation}
where $\mu$ is the cosine of the angle between the incoming and outgoing scales. Thus, the final expression for the source term in \eqref{eq:source-term-pre-final} is
\begin{equation}\label{eq:source-term-complete}
    \mathcal{S}(\bm k,t)
    =\int\frac{\dd^3\tilde{k}}{(2\pi)^{3/2}}\tilde k^2(1-\mu^2)\left(4\Phi_{\tilde{\bm{k}}}\Phi_{\bm{k}-\tilde{\bm{k}}}+2\delta\phi_{\tilde{\bm{k}}}\delta\phi_{\bm{k}-\tilde{\bm{k}}}\right).
\end{equation}


\section{Power spectrum of SIGWs}\label{sec:PS-SIGWs}

\begin{figure}
    \centering
    \includegraphics[width=\linewidth]{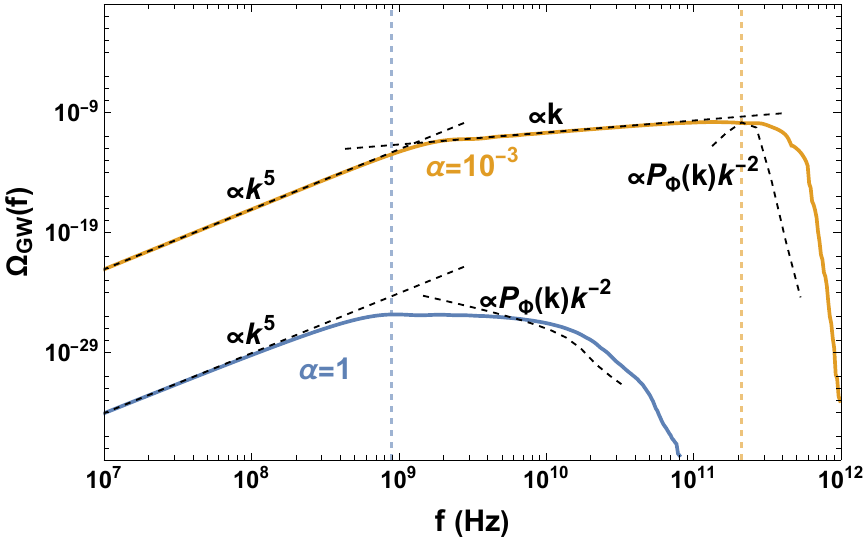}
    \caption{Fractional energy density for an E-model and two values of $\alpha$. Black dashed lines represent the analytical approximations given by eqns.~\eqref{eq:zero-k}, \eqref{eq:intermediate-k} and \eqref{eq:infinite-k}. Horizontal dashed lines mark the frequency of the peak in each case, obtained with eqn.~\eqref{eq:peak-high-alpha} for $\alpha=1$ (blue) and from the peak of $\mathcal P_{\mathcal{R}}(k)$ for $\alpha=10^{-3}$ (orange).}
    \label{fig:analytical-estimation}
\end{figure}

As shown in \cite{del-Corral:2024vcm}, the self-resonance is brief for small values of $\alpha$, followed by the usual, pressureless, matter-dominated preheating, where the scale factor evolves as $a\sim t^{2/3}$. We remark that this does not imply any fluid description since the matter content of the universe is still in the form of an oscillating scalar field. {Also after the short self-resonance phase, the perturbations $\Phi_{\bm k}$ and $\delta\phi_{\bm k}$ evolve approximately constant in amplitude} for the modes of interest. Under these assumptions, {the source term remains also constant in amplitude and therefore} $\mathcal{S}(\bm{k},t)=\mathcal S_{\bm{k}}$, and we can attempt to solve \eqref{eq:SOEOM} semi-analytically. A particular solution is given by \cite{Assadullahi:2009}:
\begin{equation}\label{eq:particular-solution}
    h_{\bm k}^{(\lambda)}(t)=\frac{\mathcal S_{\bm{k}}}{k^2}\left[1+3\,\frac{x\cos x-\sin x}{x^3}\right]=\frac{\mathcal S_{\bm{k}}}{k^2}g(k,t),
\end{equation}
where $x=\frac{2k}{aH}$, $g(k,t)$ is the growth function for the tensor modes, and we have used the following initial conditions
\begin{equation}
h_{\bm k}^{(\lambda)}(k\ll aH)=0\qquad \text{and} \qquad\dot h_{\bm k}^{(\lambda)}(k\ll aH)=0,
\end{equation}
since the typical assumption is that SIGWs are generated instantaneously when the relevant mode enters the horizon \cite{Baumann:2007}. Using \eqref{eq:particular-solution}, the two-point correlation function for $h_{\bm k}$ can be written as
\begin{widetext}
{\begin{equation}\label{eq:tensor-spectrum-pre}
\begin{split}
    \langle h_{\bm k}^{(\lambda)}(t)h_{\bm k'}^{(\lambda)}(t)\rangle&=\frac{g^2(k,t)}{k^4}\langle \mathcal S_{\bm k}\mathcal S_{\bm k'}\rangle\\
    &\simeq\frac{16g^2(k,t)}{k^4}\int\frac{\dd^3 \tilde{\bm k}}{(2\pi)^{3/2}}\int\frac{\dd^3 \tilde{\bm k}'}{(2\pi)^{3/2}}\tilde{k}^2\tilde{k}'^2(1-\mu^2)^2\bigg[\langle\Phi_{\tilde{\bm k}}\Phi_{\bm k-\tilde{\bm k}}\Phi_{\tilde{\bm k'}}\Phi_{\bm k-\tilde{\bm k'}}\rangle+\frac12\langle\delta\phi_{\tilde{\bm k}}\delta\phi_{\bm k-\tilde{\bm k}}\delta\phi_{\tilde{\bm k'}}\delta\phi_{\bm k-\tilde{\bm k'}}\rangle\bigg]\\
    &\simeq\frac{8\pi g^2(k,t)}{k^4}\delta^{(3)}(\bm k+\bm k')\int\dd^3 \tilde{\bm k}\,\frac{\tilde{k}\,(1-\mu^2)^2}{|\bm k-\tilde{\bm k}|^3}\bigg[\mathcal P_{\Phi}(\tilde k)\mathcal P_{\Phi}(|\bm k-\tilde{\bm k}|)+\frac12\mathcal P_{\delta\phi}(\tilde k)\mathcal P_{\delta\phi}(|\bm k-\tilde{\bm k}|)\bigg],
\end{split}
\end{equation}}
where in the last equality we used Wick's theorem to express the four-point correlation function of a Gaussian distribution of scalar perturbations as a function of two-point correlation functions \cite{Assadullahi:2009,Baumann:2007,Domenech:2021}. Using the definition of the power spectrum of SIGWs, eqn.~\eqref{eq:spectrum-definition}, we have 
{\begin{equation}
    \mathcal{P}^{\text{SI}}_h(k,t)=\frac{8g^2(k,t)}{\pi k}\int\dd^3 \tilde{\bm k}\,\frac{\tilde{k}\,(1-\mu^2)^2}{|\bm k-\tilde{\bm k}|^3}\bigg[\mathcal P_{\Phi}(\tilde k)\mathcal P_{\Phi}(|\bm k-\tilde{\bm k}|)+\frac12\mathcal P_{\delta\phi}(\tilde k)\mathcal P_{\delta\phi}(|\bm k-\tilde{\bm k}|)\bigg].
\end{equation}}
Finally, decomposing the volume element into spherical coordinates as
\begin{equation}\label{eq:volume-element}
    \int_V\dd^3\tilde{\bm k}=\int_0^{\infty}\tilde{k}^2\dd\tilde{k}\int_0^{\pi}\sin{\theta}\dd\theta\int_0^{2\pi}\dd\varphi=\int_0^{\infty}\tilde{k}^2\dd\tilde{k}\int_{-1}^1\dd\mu\int_0^{2\pi}\dd\varphi,
\end{equation}
we have
{\begin{equation}\label{eq:power-spectrum-appendix}
    \mathcal{P}_h^{\text{SI}}(k,t)=\frac{16g^2(k,t)}{k}\int_0^{\infty}\dd \tilde{k}\int_{-1}^1\dd\mu\,\frac{\tilde{k}^3\,(1-\mu^2)^2}{|\bm k-\tilde{\bm k}|^3}\bigg[\mathcal P_{\Phi}(\tilde k)\mathcal P_{\Phi}(|\bm k-\tilde{\bm k}|)+\frac12\mathcal P_{\delta\phi}(\tilde k)\mathcal P_{\delta\phi}(|\bm k-\tilde{\bm k}|)\bigg].
\end{equation}}
Now, to gain some insight into the behavior of the power spectrum, we analyze it in two different regimes. First, for the low-$k$ modes, we have that considering $k\ll\tilde k$ in \eqref{eq:power-spectrum-appendix}, then $|\bm k-\tilde{\bm k}|\sim \tilde k$, and thus we have
{\begin{equation}\label{eq:first-app}
    \mathcal P^{\text{SI}}_h(k,t)\sim\frac{256g^2(k,t)}{15k}\int_0^{\infty}\dd \tilde{k}\bigg[\mathcal P^{2}_{\Phi}(\tilde k)+\frac12\mathcal P^{2}_{\delta\phi}(\tilde k)\bigg]=\frac{256g^2(k,t)\,\mathcal I_{\tilde{k}}}{15k},
\end{equation}}
\end{widetext}
where $\mathcal I_{\tilde{k}}$ represents the integral of the square of the power spectra over the whole range of $\tilde k$. This essentially depends on the amplitude of the peak, so the higher the peak, the higher the integral, which ultimately translates into an amplification for the low-$k$ range, although the peak is in the high-$k$ range. This clearly reflects the convolutive nature of the source term, as stated in Sec.~\ref{sec:energy-density}.
Regarding the growth function $g(k,t)$, let us write $x=\frac{2k}{aH}$. Then, the expansion of $g(x)$ for small $x$ is $g\sim x^2/10$ and therefore, eqn.~\eqref{eq:first-app}, evaluated at the end of preheating, is given by
\begin{equation}
    \mathcal P_h^{\text{SI}}(k\ll\tilde k,t_{rh})\sim\frac{2.7\,\mathcal I_{\tilde{k}}}{k^4_{L}}k^3.
\end{equation}
Using the definition of the fractional energy density, eqn.~\eqref{eq:energy-density-final}, we have
\begin{equation}\label{eq:zero-k}
    \Omega_{\text{GW}}^{\text{SI}}(k\ll\tilde k,t_{rh})\sim\frac{0.22\Omega_{\gamma}^0\,\mathcal I_{\tilde{k}}}{k^6_{L}}k^5.
\end{equation}
For the modes higher than $k_{end}$ the growth function has settled to unity and thus $\Omega_{\text{GW}}^{\text{SI}}$ is now given by
\begin{equation}\label{eq:intermediate-k}
    \Omega_{\text{GW}}^{\text{SI}}(k\sim k_{end},t_{rh})\sim\frac{1.4\,\Omega^0_{\gamma}\,\mathcal{I}_{\tilde{k}}}{k^2_L}k.
\end{equation}
Finally, for the high-$k$ modes, we have that taking the limit $k\gg\tilde k$ in \eqref{eq:power-spectrum-appendix} we have that $|\bm k-\tilde{\bm k}|\sim k$, and the power spectrum is given by
{\begin{equation}
\begin{split}
    \mathcal P_h^{\text{SI}}(k\gg\tilde k,t_{rh})&\sim\frac{16\,\mathcal{P}_{\Phi}(k)}{k^4}\int_0^{\infty}\tilde k^3\mathcal P_{\Phi}(\tilde k)\\&+\frac{8\,\mathcal{P}_{\delta\phi}(k)}{k^4}\int_0^{\infty}\tilde k^3\mathcal P_{\delta\phi}(\tilde k)\\&=\frac{16\,\mathcal{P}_{\Phi}(k)\mathcal{I}_{k}^{\Phi}}{k^4}+\frac{8\,\mathcal{P}_{\delta\phi}(k)\mathcal{I}_{k}^{\delta\phi}}{k^4},
\end{split}
\end{equation}}
where {$\mathcal I_{k}^{\Phi}$ and $\mathcal I_{k}^{\delta\phi}$} are defined in a similar way to $\mathcal I_{\tilde{k}}$ and the growth function is set to unity. This translates into the following expression for the fractional energy density
{\begin{equation}\label{eq:infinite-k}
    \Omega_{\text{GW}}^{\text{SI}}(k\gg\tilde k,t_{rh})\sim\frac{4\Omega_{\gamma}^0\,\mathcal{P}_{\Phi}(k)\mathcal{I}_{k}^{\Phi}}{3k_U^2k^2}+\frac{2\Omega_{\gamma}^0\,\mathcal{P}_{\delta\phi}(k)\mathcal{I}_{k}^{\delta\phi}}{3k_U^2k^2}
\end{equation}}
Fig.~\ref{fig:analytical-estimation} shows these approximations as the dashed black lines for two particular values of $\alpha$ in an E-model, $\alpha=1$ (blue) and $\alpha=10^{-3}$ (orange). For the latter, we observe that \eqref{eq:intermediate-k} is useful to determine the amplitude of the peak. Considering that it is at the same scale as the peak in the power spectrum, \textit{i.e.} $k_p$, then the maximum amplitude of the fractional energy density is computed as
\begin{equation}\label{eq:Omega-max}
    \Omega_{GW}^{\text{SI},\text{max}}\sim \frac{1.4 \Omega_{\gamma}^0\mathcal I_{\tilde{k}}}{k_L^2}k_p,
\end{equation}
where $k_p\sim k_U/\sqrt{2}$ \cite{Hertzberg:2014iza,Hertzberg:2014jza}. Eqn.~\eqref{eq:Omega-max} is shown as the vertical blue dashed line of Fig.~\ref{fig:analytical-estimation}. On the contrary, for $\alpha=1$, since there is no amplification peak, the maximum is at the point where the growth function peaks. Again, considering $x=\frac{2k}{aH}$, these points are given by the roots of the equation
\begin{equation}\label{eq:peak-high-alpha}
    \tan{x}=\frac{3x}{3-x^2},
\end{equation}
the first of which is given approximately by $x_0\sim5.76$, which implies $k_0\sim2.88k_L$. This is shown as the vertical orange dashed line of Fig.~\ref{fig:analytical-estimation}. Finally, for comparison purposes, we also show the fractional energy density of BGWs for the same values of $\alpha$ but in lighter colors. One can observe that, for the Starobinsky case, the background contribution is higher than the induced one, contrary to the case of $\alpha=10^{-3}$.

\section{Comments on the results from Lattice simulations}

The amplitude of scalar-induced gravitational waves (SIGWs) during the preheating regime has been the subject of investigations in the literature \cite{Zhou:2013tsa,Lozanov:2019ylm,Antusch:2016con}. We cannot make a fair comparison of our results with these approaches because of structural and methodological differences. 

This discrepancy can be traced to methodological differences between the lattice-based “oscillon” approach and our perturbative treatment, particularly regarding the treatment of metric fluctuations for modes with $k > aH$. Below, we summarize the main aspects of the oscillon-based calculations \cite{Hasegawa:2017iay,Shafi:2024jig,Antusch:2016con} and clarify the assumptions underlying each framework.
Historically, oscillons were identified as localized, long-lived solutions of scalar field equations in Minkowski spacetime \cite{Copeland:1995fq}. Modern lattice implementations such as \texttt{CosmoLattice} \cite{Figueroa:2021yhd,Caravano:2021pgc} (which are improved versions of LatticeEASY \cite{Felder:2000hq}, DEFROST \cite{Frolov_2008}, and CUDAEasy \cite{Sainio:2009hm}) build on this framework and study the nonlinear evolution of scalar fields in an expanding universe. These simulations are extremely useful for investigating nonperturbative scalar dynamics, including oscillon formation and decay. However, they generally evolve the scalar field on a homogeneous background metric, with the expansion rate determined by a spatially averaged energy density.

More concretely, the equations solved in such approaches are 
\begin{equation}
\begin{aligned}
\ddot{\phi}+3H\dot{\phi}-\frac{\nabla^2\phi}{a^2}+ V_{,\phi}& =0,\\
H^2 & = \frac{8\pi G}{3}\Big\langle \frac{\dot{\phi}^2}{2}+\frac{(\nabla\phi)^2}{2a^2}+V(\phi)\Big\rangle,
\end{aligned}
\label{Lateq}
\end{equation}
where $\langle\cdots\rangle$ denotes a spatial average. This prescription effectively replaces the local Einstein equations,
\begin{equation}
G_{\mu\nu}[g_{\mu\nu}(t,\bm x)] = 8\pi G,T_{\mu\nu}[\phi(t,\bm x)],
\end{equation}
with a single background equation sourced by the averaged energy density. While this provides a tractable description of the mean expansion, it does not capture the local metric fluctuations $\delta g_{\mu\nu}(t,\bm x)$ that are sourced by inhomogeneities in $\phi(t,\bm x)$. As a result, the effects of scalar, vector, and tensor perturbations, including scalar-induced gravitational waves, are not fully represented within this framework. In our approach, by contrast, we solve the linearized Einstein equations consistently for both the scalar field and the metric perturbations. This allows us to track the amplification of scalar fluctuations and the subsequent generation of SIGWs in a self-consistent manner. Because of these conceptual differences, direct quantitative comparison between our results and those of Refs.~\cite{Bhoonah:2020oov,Hasegawa:2017iay,Shafi:2024jig} is not straightforward. Nevertheless, it is worth emphasizing that the two approaches are complementary. Lattice simulations offer valuable insight into the nonlinear scalar dynamics and oscillon formation, while our treatment focuses on the interplay between scalar perturbations and metric fluctuations in the linear regime. A fully nonlinear numerical framework that consistently includes metric perturbations could provide a natural bridge between these two perspectives.

\begin{figure}
    \includegraphics[width=\textwidth]{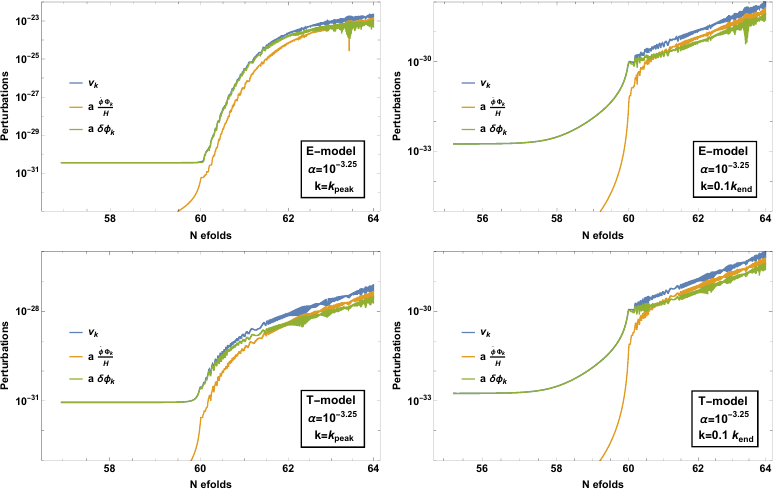}
    \caption{Time evolution of the components of the MS variable, Eqn.~\eqref{eq:MSvariable} in their natural units {(i.e., multiplied by the factors of $k^{3/2}$)}. The mode $k_{\text{peak}}$ refers to the wavenumber whose perturbations are amplified the most. That is, the peak of the power spectrum. The wavenumber $k=0.1k_{\rm end}$ chosen refers to a mode that exits the horizon during inflation and then re-enters during preheating. }
    \label{fig:MScomponents}
\end{figure}

{To further illustrate the source of discrepancies between the lattice-based “oscillon” approach and our perturbative treatment, it is worth keeping track, in the $\alpha-$attractor case, of the behavior of the metric perturbations in the transition between the end of inflation and the preheating stage. In Fig.~\ref{fig:MScomponents}, we separately show the evolution of the scalar-field and metric contributions (green and brown lines, respectively) to the Mukhanov-Sasaki variable $v_k$ (blue lines). Before the start of preheating, the metric contribution is suppressed, as it is proportional to $\dot{\bar\phi}/(\Mpl H) =\sqrt{2\epsilon_H}\ll 1$ during slow roll. At the onset of preheating, however, the metric contribution is no longer suppressed, and becomes of the same order as the contribution from the scalar field. Thus the metric fluctuations can clearly not be ignored during the preheating instability. This can be easily understood from the \eqref{Phidelphi} with an approximation $\dot\Phi \approx 0$ (which is true for the perturbation inside the instability band once it gets amplified due to preheating instability; see~Fig.~\ref{fig:scheme} for example) the metric fluctuation becomes of the same order as the scalar field fluctuation (since $\dot{\bar{\phi}}/(\Mpl H)=\sqrt{2\epsilon_H}\sim \mathcal{O}(1)$ during preheating)
\begin{equation}
    \Phi_\textbf{k} \approx \frac{1}{2\Mpl^2} \frac{\dot{\bar{\phi}}}{H} \delta\phi_\textbf{k}
\end{equation}
This behavior is general and applies beyond the specific $\alpha$-attractor models we discuss.}

It is worth noticing that while \cite{Bhoonah:2020oov} claimed to include the metric fluctuations in the evaluation of SIGW in $\alpha$-attractor models, their simulation is based on the HLattice code \cite{Huang:2011gf}, which, as stated explicitly after their Eq.~26, only includes contributions up to second-order in the perturbations of the Einstein-Hilbert term. On the contrary, we explicitly include contributions of 3rd order in the perturbations of the action \eqref{3rdordac}.

{Therefore, since all the lattice-based studies rely on spatial averaging of the scalar field while completely or partially ignoring the metric fluctuations, it is not trivial to make any detailed comparison with our work.}

\onecolumngrid

\twocolumngrid


\bibliographystyle{apsrev4-2}
\bibliography{BIBLIO.bib}

\end{document}